\documentclass[12pt,a4paper,final]{iopart}             
\usepackage{iopams,comment}							   
\usepackage{amsthm}									   
\usepackage[dvips]{graphicx,epsfig}					   
\usepackage{pst-plot}								   
\usepackage[breaklinks=true,colorlinks=true,		   
            linkcolor=black,urlcolor=black,			   
            citecolor=black]{hyperref}				   

\begin{document}

\title[Information-theoretic measures for a PDM system
in an infinite potential well]{Information-theoretic measures for a
position-dependent mass system in an infinite potential well}

\author{Bruno G da Costa$^1$ and Ignacio S Gomez$^2$}
\address{$^1$ Instituto Federal de Educa\c{c}\~ao, Ci\^encia e
Tecnologia do Sert\~ao Pernambucano,
Rua Maria Luiza de Ara\'ujo Gomes Cabral s/n, 56316-686
Petrolina, Pernambuco, Brazil}
\address{$^2$ Instituto de F\'isica, Universidade Federal da Bahia,
Rua Bar\~ao de Jeremoabo s/n, 40170-115 Salvador, Bahia, Brazil}
\eads{\mailto{bruno.costa@ifsertao-pe.edu.br},
\mailto{nachosky@fisica.unlp.edu.ar}}

\vspace{10pt}
\begin{indented}
\item[] \today
\end{indented}

\begin{abstract}
In this work we calculate the Cram\'{e}r--Rao, the Fisher--Shannon
and the L\'{o}pez-Ruiz--Mancini--Calbert (LMC) complexity measures
for eigenstates of a deformed Schr\"{o}dinger equation, being
this intrinsically linked with position-dependent mass (PDM) systems.
The formalism presented is illustrated with a particle confined in
an infinite potential well.
Abrupt variation of the complexity near to the asymptotic value
of the PDM-function $m(x)$
and erasure of its asymmetry along with
negative values of the entropy density in the position space,
are reported as a consequence of the interplay between the
deformation and the complexity.
\end{abstract}

\vspace{2pc}

\section{\label{sec:intro}
         Introduction}

The complexity of a system is understood as a measure of internal
order/disorder of its components and of the interrelations between
them.
From the standard description of considering the states of a system
characterised by probability density distributions, several notions
of complexity can be defined in terms of statistical measures as
the information entropy, within the context of the information theory
\cite{Shannon-book-1949,Fisher-1925,
      Frieden-book-2004,Cover-book-2006}.
Intuitively, a fully ordered system (for instance a crystal)
does not present complexity or its complexity is zero.
The same simple description can be assigned to a totally disordered
system like an ideal gas, although their physical characteristics
differ a lot from a crystal.
In these basic examples we see that they share in common a minimal
level of complexity. Several measures of complexity for
finite systems in terms of the variance, the Shannon entropy
\cite{Shannon-book-1949}, the Fisher information \cite{Fisher-1925}
and other statistical measures have been proposed
\cite{Gell-Mann-1996,
      Catalan-2002,
      Yamano-2004,
      Romera-2011,
      Sen-book-2012,
      Chamon-2013,
      Sanchez-Moreno-2014,
      Merhav-2015,
      Sobrino-Coll-2017}.
In this context, two of the most popular are
the Cram\'{e}r-Rao \cite{Dembo-1991,Dehesa-2006}
and the Fisher-Shannon \cite{Angulo-2008,Romera-2004}
complexities, which are monotones with respect to
a convolution with any Gaussian probability distribution, i.e., the
complexity after the convolution decreases with respect to the
original one.
In addition, the LMC (L\'{o}pez-Ruiz, Mancini and Calbert
\cite{LMC-1995,Piqueira-2011})
complexity is monotonic regarding to a certain family of
stochastic operations. Subsequently, quantum information measures
and their properties against LOCC
operations have motivated the following axiomatic definition of
complexity \cite{Rudnicki-2016}:
a preexistent family of probability density
distributions of minimal complexity
that are preserved by a certain class of operations, and the
monotonicity of the complexity respect
to these operations.

Furthermore,
along several decades an increasing application
of these three measures of complexity (the Cr\'{a}mer-Rao,
the Fisher-Shannon, and the LMC) have been reported
in order to characterise a wide class of phenomena and systems:
Hartree-Fock wave functions \cite{Angulo-2008},
multiparticle systems \cite{Romera-2004},
chemical reactions \cite{Angulo-2008-Wiley},
Morse and P\"oschl--Teller potentials \cite{Dehesa-2006-PT},
rigid rotator \cite{Dehesa-2015},
hydrogenic-like systems
\cite{Yanez-1994,Guerrero-2011,Nascimento-Prudente-2018},
blackbody radiation problem \cite{Toranzo-Dehesa-2014},
Dirac-delta-like quantum potentials \cite{Bouvrie-Angulo-2011},
particle in an infinite well \cite{Lopez-Rosa-2011},
etc.
Among these systems, those which have a position-dependent mass
(i.e., the so-calles position-dependent mass systems,
or briefly PDM systems)
have presented a particular interest
due to applicability in multiple areas:
astrophysics \cite{Richstone_1982},
semiconductors \cite{vonroos_1983},
quantum dots \cite{Serra-Lipparini-1997},
quantum liquids \cite{Barranco-1997},
inversion potential for ${\rm NH}_3$ \cite{Aquino_1998},
many body theory \cite{Bencheikh-et-al-2004},
relativistic quantum mechanics \cite{Alhaidari-2004},
superintegrable systems \cite{Ranada-2016},
supersymmetry \cite{Bravo-PRD-2016},
nuclear physics \cite{Alimohammadi-Hassanabadi-Zare-2017},
nonlinear optics \cite{Li-Guo-Jiang-Hu},
etc.
In addition, studies of information-theoretic measures for PDM
systems have been made in references
\cite{YanezNavarro-2014,Dong-2014,Guo-Hua-2015,
Falaye-2016,Serrano-2016, Macedo-Guedes-2015}.

One of the main achievements of the PDM systems
is that they allow to model the dynamics of particles in an
non-homogeneous media.
Lately, a deformed Schr\"{o}dinger equation associated
with a position-dependent mass has been proposed
within the context of generalised translation operators
\cite{CostaFilho-Almeida-Farias-AndradeJr-2011,
CostaFilho-Alencar-Skagerstam-AndradeJr-2013,
Mazharimousavi-2012,
Vubangsi-Tchoffo-Fai-2014,Costa-Borges-2014,
Costa-Borges-2018,Costa-Gomez-2018,
Nobre-2017,Souza-2019,Chung-2019,
Gomez-RPM-2019}, where a real parameter
$\gamma$ controls the deformation and whose mathematical
background is provided by an algebraic structure
\cite{Nivanen_2003,Borges_2004}
inspired in nonextensive statistics
\cite{Tsallis-book-2009,Entropia-Tsallis}.

Motivated by previous works
\cite{CostaFilho-Almeida-Farias-AndradeJr-2011,
CostaFilho-Alencar-Skagerstam-AndradeJr-2013,
Costa-Borges-2014,Costa-Borges-2018,
Costa-Gomez-2018,Gomez-RPM-2019}
and following the definition of complexity given in
\cite{Rudnicki-2016}, in this work we calculate the Cram\'{e}r-Rao,
the Fisher-Shannon and the LMC complexities for eigenstates
of the deformed Schr\"{o}dinger equation
for a particle in an infinite potential well
\cite{CostaFilho-Almeida-Farias-AndradeJr-2011,
Mazharimousavi-2012,Costa-Borges-2014}
and then, we analise the interplay between the
underlaying deformed structure and the behavior of the complexities.
In particular, we extent the results presented in
\cite{Lopez-Rosa-2011} for a particle with PDM.

The paper is organised as follows.
In Section 2 we review the notion of complexity
for probability density distributions
as well as the entropic uncertainty relation.
Section 3 is devoted to calculate the
Cram\'{e}r-Rao, the Fisher-Shannon and the LMC complexities
for eigenstates of the deformed Schr\"{o}dinger of a
particle confined in an infinite potential well.
Finally, in Section 4 we outline some conclusions
and future directions are discussed.

\section{\label{sec:preliminaries}
         Preliminaries}

In this Section we give the necessary concepts and definitions
for the development of the forthcoming sections.

\subsection{\label{subsec:complexities}
		   Complexity measures for probability density distributions}

We introduce some basic elements of the information-theoretic
description of unidimensional probability density distributions.
We consider continuous probability density distributions
$\rho(x)$ defined over a subset $X$ of the real numbers $\mathbb{R}$,
mathematically defined by
$\rho: \mathbb{R}\rightarrow [0,+\infty)$
with $\int_X \rho(x)\rmd x=1$.
The spread of the variable $x$ over an interval
$X \subseteq\mathbb{R}$
is measured by its standard deviation
or \emph{Heisenberg length} given by the
square mean deviation value of $x$
\begin{equation}
\label{eq:lentgh-Heis}
L_{{\rm H}}=\Delta x
=\sqrt{V[\rho]}=\sqrt{\langle x^2\rangle - \langle x \rangle^2}
\end{equation}
where $V[\rho]$ is the variance of $x$.
As usual, $\langle f(x) \rangle$ stands for the mean value of $f(x)$
with respect of the probability density distribution $\rho(x)$,
i.e., $\langle f(x)\rangle=\int_{X}f(x)\rho(x)\rmd x$.
Given a probability density distribution $\rho$, the Shannon
entropy (or information entropy) $S[\rho]$ and the Fisher information
$F[\rho]$ are found between the most relevant information theoretic
measures. The former, introduced by
C. Shannon in his foundational work on communication theory
\cite{Shannon-book-1949}, is the expected value of the information
entropy of the distribution $\rho(x)$.
The latter, whose role in estimation theory was pointed out by
R. Fisher \cite{Fisher-1925},
measures the information of the variable $x$ regarding
an unknown parameter associated to the distribution $\rho(x)$.
These are given by
\begin{equation}
\label{eq:Shannon-entropy}
S[{\rho}] = - {\int}_{\! \! X}  \, \rho(x) \ln {\rho}(x)\rmd x
      	  = - \langle \ln \rho (x) \rangle
\end{equation}
and
\begin{equation}
\label{eq:Fisher-information}
F[{\rho}] = {\int}_{\! \! X} \,	
		\frac{1}{{\rho}(x)}
		\left[ \frac{\rmd \rho (x)}{\rmd x} \right]^2 \rmd x
		= \left\langle \left[
          \frac{\rmd \ln \rho (x)}{\rmd x}
          \right]^2 \right\rangle
\end{equation}
respectively. Other information measure of interest
is the Onicescu entropy or disequilibrium \cite{Onicescu}:
\begin{equation}
\label{eq:disequilibrium}
	D[{\rho}] =
	{\int}_{\! \! X} \,	
	[{\rho}(x)]^2 \rmd x
	= \langle {\rho}(x) \rangle,
\end{equation}
which corresponds to the entropic moment of second order.
The Shannon information and the disequilibrium are global
quantifiers that measure the extent to which the density is
concentrated, while the Fisher information is local
(due to the derivative of $\rho$ contained in its definition)
and it characterises the oscillatory nature of $\rho$.
These measures complement each other and they characterise
different aspects of the distribution $\rho$.
For comparing, the Shannon and Fisher lengths
(denoted by $L_{{\rm S}}$ and $L_{{\rm F}}$)
are defined \cite{Bouvrie-Angulo-2011,Toranzo-Dehesa-2014}
\begin{equation}
\label{eq:Shannon-Fisher-length}
L_{{\rm S}}= \rme^{S[\rho]}, \qquad
L_{{\rm F}}= \frac{1}{\sqrt{F[\rho]}}.
\end{equation}
deserve to be mentioned.
Heisenberg length measures the separation between the regions,
where the probability density is concentrated, regarding the mean
value $\langle x \rangle$. Shannon length quantifies the
concentration of the density distribution $\rho$.
Fisher length tends to zero for discontinuous distributions
and it presents a high sensibility to fluctuations of $\rho$.
Moreover, we have the following relations:
\begin{equation}
\label{eq:relations-lengths}
L_{{\rm F}}\leq L_{{\rm H}} \quad \ , \ \quad
(2\pi e)^{\frac{1}{2}}L_{{\rm F}}
\leq L_{{\rm S}}\leq
(2\pi e)^{\frac{1}{2}}L_{{\rm H}}
\end{equation}
where the equality is satisfied for the Gaussian distributions.
It is worth to be mentioned that other notions of the
information length have been extended, for instance in the context of
the Lagrangian formulation \cite{Kim-2018}.
The complexity measures associated to the information measures
$S[\rho], F[\rho], D[\rho]$
with their respectives Heisenberg, Shannon and Fisher lengths
are constructed by means of the product between
the information measure and its corresponding length,
thus capturing a joint balance of the features of $\rho$ described
by the information measures and the lengths.
In this way, the Cram\'{e}r-Rao, Fisher-Shannon and LMC
complexity measures are defined as \cite{Rudnicki-2016}:
\numparts
\begin{eqnarray}
\label{eq:CR-complexity}
C_{{\rm CR}}[{\rho}] &=&
F[{\rho}](\Delta x)^2
= F[\rho] \times L_{{\rm H}}^2, \\
\label{eq:FS-complexity}
C_{{\rm FS}}[\rho] &=&
\frac{1}{2\pi e} F[\rho] \times  \rme^{2S[{\rho}]}
= \frac{1}{2\pi e}F[\rho] \times L_{{\rm S}}^2,\\
\label{eq:LMC-complexity}
C_{{\rm LMC}}[\rho] &=&
D[{\rho}] \times \rme^{S[{\rho}]}
= D[\rho] \times L_{{\rm S}}.
\end{eqnarray}
\endnumparts
It should be noted that these complexity measures
are (i) dimensionless, (ii) lower bounded
by the unity, (iii) minimum for the Dirac delta (maximum order) and
the uniform distribution (maximum disorder), and (iv)
invariant under translation and scaling transformations.

\subsection{Entropic uncertainty relations}

Entropic uncertainty relations are important expressions that allow
to characterise several dynamical features of a quantum system
concerning two non-commuting observables. Generally, for two
non-commuting observables $\hat{A}$ and $\hat{B}$ an entropic
uncertainty relation presents the form \cite{BBM-1975,Yanez-1994}
\begin{equation}
\label{eq:entropic-uncertainty}
S_{\hat{A}}+S_{\hat{B}}\geq
h_{\hat{A}\hat{B}}
\end{equation}
where $S$ is a functional defined over the set
of observables of the quantum system (typically the entropy),
and $h_{\hat{A}\hat{B}}$ is a positive constant that contains
information about the non-commutativity between the observables
$\hat{A},\hat{B}$ and it represents the lower bound of the
uncertainty relation.
As an example, let us consider $\hat{A}=\hat{x}$ and
$\hat{B}=\hat{k}$ the position and wave-vector operators,
$\psi(x)$ the wave-function in the position space,
$\widetilde{\psi}(k)$ the wave-function in the $k$-space,
and $\mathcal{S}$ a modified
Shannon entropy for the distributions $\rho(x)=|\psi(x)|^2$
and $\widetilde{\rho}(k) = |\widetilde{\psi}(k)|^2$
\cite{Nascimento-Prudente-2018},
i.e.,
\numparts
\begin{equation}
\label{eq:S_x}
\mathcal{S}_x
[{\rho}]
	= -\int_{-\infty}^{\infty} {\rho}(x)
		\ln [{\sigma \rho}(x)]\rmd x
	= - \langle \ln [ \sigma \rho (x)] \rangle
	= S[{\rho}] - \ln \sigma
\end{equation}
and
\begin{equation}
\label{eq:S_k}
	\mathcal{S}_k[\widetilde{\rho}]
	= -\int_{-\infty}^{\infty} {\widetilde{\rho}}(k)
		\ln \left[{\widetilde{\rho}} (k)/\sigma \right] \rmd k
	= - \langle \ln [\widetilde{\rho} (k)/ \sigma ] \rangle
	= S[\widetilde{\rho}] + \ln \sigma,
\end{equation}
\endnumparts
where $\sigma$ is introduced in order
to have a dimensionless argument of the logarithm.
Then, using the equations (\ref{eq:S_x}) and
(\ref{eq:S_k})
the entropic uncertainty relation (\ref{eq:entropic-uncertainty})
adpots the BBM-inequality form \cite{BBM-1975}
\begin{equation}
\label{eq:entropic-uncertainty-x-k}
\mathcal{S}_x[{\rho}] + \mathcal{S}_k[\widetilde{\rho}]
= S[{\rho}] + S[\widetilde{\rho}] \geq 1 + \ln \pi.
\end{equation}
%

\section{\label{sec:complexity}
         Complexity measures for a particle with a position-dependent
         effective mass in an infinite potential well}

In this Section we calculate the Cram\'{e}r-Rao, the Fisher-Shannon
and the LMC complexities (given by
(\ref{eq:CR-complexity}), (\ref{eq:FS-complexity}) and (\ref
{eq:LMC-complexity})) as well as the entropic uncertainty relations
for the eigenstates of a unidimensional particle with a PDM and
confined in an infinite potential well within the displacement
operator approach
\cite{CostaFilho-Almeida-Farias-AndradeJr-2011,
      CostaFilho-Alencar-Skagerstam-AndradeJr-2013,
      Mazharimousavi-2012,Vubangsi-Tchoffo-Fai-2014,
      Costa-Borges-2014,Costa-Borges-2018,Costa-Gomez-2018}.

\subsection{\label{subsec:q-schrodinger}
		Schr\"odinger equation for position-dependent effective mass
		and deformed space}

Costa Filho \etal have introduced a quantum system with a PDM by
using a generalised translation operator which produces nonadditive
spatial displacements expressed by
\cite{CostaFilho-Almeida-Farias-AndradeJr-2011}
\begin{equation}
\label{eq:T_gamma}
	\hat{\tau}_{\gamma}(\varepsilon)|x \rangle
			= | x + \varepsilon + \gamma x \varepsilon \rangle,
\end{equation}
with $\varepsilon$ an infinitesimal displacement and $\gamma$
a parameter with dimensions of inverse length.
The operator $\hat{\tau}_{\gamma}(\varepsilon)$
sends a well-localised state around $x$ into another well-localised
state around $x+\varepsilon +\gamma x\varepsilon$,
without changing the other physical properties.
In this sense, the parameter $\gamma$ can be understood
as a measure of coupling between the particle displacement
$\varepsilon$ and the original position $x$.
The generator of translations (\ref{eq:T_gamma}) is
the Hermitian momentum operator
\cite{Mazharimousavi-2012,Vubangsi-Tchoffo-Fai-2014,
Costa-Borges-2014}
\begin{eqnarray}
\label{eq:Pi-momentum}
\hat{\Pi} &= \frac{(\hat{1} + \gamma \hat{x}) \hat{p}}{2}
  		     + \frac{\hat{p}(\hat{1} + \gamma \hat{x})}{2}
			 \nonumber \\
		&= (\hat{1} + \gamma \hat{x})^{1/2} \hat{p}
				  (\hat{1} + \gamma \hat{x})^{1/2}.
\end{eqnarray}
The canonically conjugate space operator
for the deformed linear momentum $\hat{\Pi}$ is
\begin{equation}
\label{eq:eta}
	\hat{\eta} = \frac{\ln ( \hat{1} + \gamma \hat{x})}{\gamma}.
\end{equation}
Thus, $(\hat{\eta}, \hat{\Pi})\rightarrow (\hat{x}, \hat{p})$
results a point canonical transformation (PCT) which maps
a particle with constant mass
$m_0$ into other with position-dependent mass.
In fact, the Hamiltonian operator
$
\hat{K}(\hat{\eta}, \hat{\Pi})
	= \frac{1}{2m_0}\hat{\Pi}^2 + \hat{U}(\hat{\eta})
$
is mapped into
$\hat{H} (\hat{x}, \hat{p})= \hat{T} +\hat{V}(\hat{x})$
with
$\hat{V}(\hat{x}) = \hat{U}(\hat{\eta}(\hat{x}))$
the potential energy operator,
\begin{equation}
\label{eq:general-hamiltonian-pdm}
	\hat{T} =
    \frac{1}{2}[m(\hat{x})]^{-1/4}\hat{p}\,[m(\hat{x})]^{-1/2}
    \hat{p}\,[m(\hat{x})]^{-1/4},
\end{equation}
the kinetic energy operator, and
\begin{equation}
\label{eq:m(x)}
m(x) = \frac{m_0}{(1 + \gamma x)^2}
\end{equation}
the effective mass. This function has a singularity in
$x_{\rm d} = -1/\gamma$ due to the definition of the deformed
translation operator (\ref{eq:T_gamma}) which satisfies the property
$\hat{\mathcal{T}}_{\gamma}(\varepsilon)|x_{{\rm d}} \rangle
			= | x_{{\rm d}} \rangle$,
i.e., the particle has an inertia so high in $x_{\rmd}$
that it can not be displaced whatever the value of $\varepsilon$.
Since we are interested in probability density distributions
given by stationary wave functions then we only consider the
time-independent Schr\"{o}dinger equation. Consequently, from
(\ref{eq:general-hamiltonian-pdm}) and (\ref{eq:m(x)})
the time-independent Schr\"odinger equation
$\hat{H}|\alpha \rangle = E|\alpha \rangle$
in terms of the wave function
$\psi(x) = \langle x | \alpha \rangle$
results
\begin{eqnarray}
\label{eq:schrodinger-equation-sho-pdm}
&-\frac{{\hbar}^2 ( 1+\gamma x)^2}{2m_0}
\frac{\rmd ^2 \psi (x)}{\rmd x^2}
-\frac{{\hbar}^2\gamma( 1+\gamma x)}{m_0}\frac{\rmd \psi (x)}{\rmd x}
-\frac{{\hbar}^2\gamma^2}{8m_0}\psi (x)
+ V(x) \psi (x)
\nonumber \\
&= E \psi (x).
\end{eqnarray}
\noindent
Alternatively, the equation (\ref{eq:schrodinger-equation-sho-pdm})
can be expressed by means of a field $\varphi(x)$ related to
$\psi (x)$ by \cite{Costa-Borges-2018,Costa-Gomez-2018}
\begin{equation}
\label{eq:psi-phi-relation}
	\psi (x) = \sqrt[4]{\frac{m(x)}{m_0}} \varphi (x)
				= \frac{\varphi (x)}{\sqrt{1+\gamma x}}.
\end{equation}
for $x > -1/\gamma$ and $\psi (x) = 0$ otherwise,
in such a way that the region $(-\infty, -1/\gamma)$
is forbidden for the particle with the mass function
given by (\ref{eq:m(x)}).
Therefore $\rho (x) = |\psi(x)|^2 = |\varphi(x)|^2/(1+\gamma x)$
remains nonnegative $\forall x$, in accordance with the concept
of probability density.
Substituting (\ref{eq:psi-phi-relation}) in
(\ref{eq:schrodinger-equation-sho-pdm}), one obtains
\begin{equation}
\label{eq:deformed-schrodinger-equation-sho-pdm}
-\frac{{\hbar}^2}{2m_0} \hat{D}_\gamma^2 \varphi (x) + V(x) \psi (x)
= E \varphi (x),
\end{equation}
where $\hat{D}_\gamma = (1+\gamma x)\frac{\rmd}{\rmd x}$ is a deformed derivative operator
\cite{Costa-Borges-2014,Costa-Borges-2018,
      Borges_2004,Costa-Gomez-2018}.
By means of the change of variable
$x\rightarrow \eta =\gamma^{-1} \ln (1+\gamma x)$,
we recover the time-independent Schr\"odinger equation
for a particle with constant mass $m_0$ in the
representation of the deformed space
$\{ |  \eta \rangle \}$
\cite{CostaFilho-Almeida-Farias-AndradeJr-2011,
      CostaFilho-Alencar-Skagerstam-AndradeJr-2013,
      Costa-Borges-2018}:
\begin{equation}
\label{eq:schrodinger-equation-deformed space}
-\frac{\hbar^2}{2m_0} \frac{\rmd ^2 \phi ({\eta})}{\rmd  \eta^2}
  	+ U({\eta})\phi ({\eta}) = E\phi ({\eta}),
\end{equation}
described in terms of the wave function
$\phi({\eta}) = \varphi (x({\eta}))$.

The eigenfunctions of the momentum operator (\ref{eq:Pi-momentum})
in the $x$-representation, with
$\hat{\Pi}|k\rangle=\hbar k|k\rangle$, are \cite{Costa-Borges-2014}
\begin{eqnarray}
\label{eq:eigeinfunctions-p-x-representation}
	\psi_k (x) = \langle x | k \rangle
	           = \frac{\psi_0}{\sqrt{1+ \gamma x}}
			     \exp \left[\frac{\rmi k}{\gamma} \ln (1+\gamma x)
                 \right],
\end{eqnarray}
which are non-normalizable and constitute a deformation of the
standard free particle plane waves, being this latter recovered for
$\gamma \rightarrow 0$.
A wave packet for position-dependent mass system can be defined by
the deformed Fourier transform:
\begin{equation}
\label{eq:wave-packet-pdm-psi}
	\psi (x) = \frac{1}{\sqrt{2\pi}} \frac{1}{\sqrt{1 + \gamma x}}
			   \int_{-\infty}^{+\infty}
			   \widetilde{\psi} (k) \rme^{\rmi k\gamma^{-1} \ln (1 + \gamma x)}\rmd k,
\end{equation}
with $\widetilde{\psi}(k)$ a distribution function of the
wave vectors $k$.
From relation (\ref{eq:psi-phi-relation})
and the coordinate transformation $x \rightarrow \eta$,
the deformed inverse Fourier transform of (\ref{eq:wave-packet-pdm-psi}) results
\begin{eqnarray}
\label{eq:inverse-deformed-transform}
\widetilde{\psi} (k)
            &= \frac{1}{\sqrt{2\pi}} \int_{-\infty}^{+\infty}
            \phi ({\eta}) \rme^{-\rmi k{\eta}}\rmd \eta
			\nonumber \\
            &= \frac{1}{\sqrt{2\pi}} \int_{-\infty}^{+\infty}
			   \frac{\varphi (x)}{1+\gamma x}
			   \rme^{-\rmi k\gamma^{-1} \ln (1 + \gamma x)} \rmd x
			\nonumber \\
            &= \frac{1}{\sqrt{2\pi}} \int_{-\infty}^{+\infty}
			   \frac{\psi (x)}{\sqrt{1+\gamma x}}
			   \rme^{-\rmi k\gamma^{-1} \ln (1 + \gamma x)} \rmd x.
\end{eqnarray}

\subsection{Particle in an infinite potential well}

Consider a unidimensional particle with a PDM $m(x)$ given by
(\ref{eq:m(x)}) in an infinite one-dimensional square potential well
of width $2a$:
\begin{equation}
V(x) = \left\{
\begin{array}{ll}		
\displaystyle
0 		& -a < x < a, \\
\infty, &  {\rm otherwise}.
\end{array}
\right.
\end{equation}
Outside the square potential well the solution
$\psi(x)$
is null outside
because this region is forbidden for the particle.
Inside,
the solution $\psi(x)$ of the Schr\"odinger equation
(\ref{eq:schrodinger-equation-sho-pdm}) is
\begin{equation}
\label{eq:autofunction-generalized}
	\psi(x) = \frac{1}{\sqrt{1+ \gamma x}}
				\left[
				C_{+}
				e^{\rmi k \gamma^{-1} \ln (1+\gamma x)}
				+ C_{-}
				e^{-\rmi k \gamma^{-1} \ln (1+\gamma x)}
				\right]
\end{equation}
where $k=\frac{\sqrt{2m_0 E}}{\hbar}$ and $C_{\pm}$ are constants.
From the boundary conditions $\psi(a) = \psi(-a) = 0$, we obtain
\begin{equation}
	\left(
	\begin{array}{ll}		
	\displaystyle
	\rme^{\rmi k{\gamma}^{-1}\ln (1 + \gamma a)} &
	\rme^{-\rmi k{\gamma}^{-1}\ln (1 + \gamma a)} \\
	\rme^{\rmi k{\gamma}^{-1}\ln (1 - \gamma a)} &
	\rme^{-\rmi k{\gamma}^{-1}\ln (1 - \gamma a)}
	\end{array}
	\right)
	\left(
	\begin{array}{c}		
	\displaystyle
	C_{+} \\ C_{-}
	\end{array}
	\right)
	= \left(
	\begin{array}{c}		
	\displaystyle
	0 \\ 0
	\end{array}
	\right).
\end{equation}
This equation has nontrivial solutions only if the determinant
of the $2 \times 2$ matrix vanishes, which leads us to
$
\sin \left[
\frac{k}{\gamma} \ln \left( \frac{1+\gamma a}{1+\gamma a} \right)
\right] = 0,
$
that is $k_{\gamma,n} = n \pi/L_\gamma$ with $n$ an integer
\begin{equation}
\label{eq:Lgamma}
L_\gamma = \frac{1}{\gamma}
\ln \left(\frac{1 + \gamma a}{1 - \gamma a}\right)
= 2a\left[ \frac{{\rm atanh} (\gamma a)}{\gamma a}\right].
\end{equation}
$L_\gamma$ is the length of box at the
coordinate basis $\{ | \eta \rangle \}$ which
can also be obtained by the transformation
$x \rightarrow \eta$.
Therefore, the eigenfunctions for this problem are
\begin{equation}
\label{eq:psi_n}
	\psi_n (x) = \frac{A_\gamma}{\sqrt{1 + \gamma x}}
			     \sin \left[ \frac{k_{\gamma,n}}{\gamma}
                 \ln \left( \frac{1 + \gamma x}{1 + \gamma a} \right) \right]
\end{equation}
at the interval $|x| < a$  and $\psi_n(x) = 0$ otherwise with
$A_{\gamma} = \sqrt{2/L_\gamma}$ the normalisation constant.
The energy levels are given by
\begin{equation}
\label{eq:energy-box-pdm}
	E_n = \frac{\hbar^2 \pi^2 n^2}{2m_0 L_\gamma^2}
		=\varepsilon_0 n^2
		  \left[ \frac{\gamma a}{{\rm atanh} (\gamma a)} \right]^2,
\end{equation}
with $\varepsilon_0 = \hbar^2 \pi^2/8m_0 a^2$.
The probability densities of the stationary states in the
position-space are
\begin{equation}
\label{eq:rho_n(x)}
\rho_n (x) = |\psi_n (x)|^2 =
\frac{2\gamma}{\ln \left( \frac{1+\gamma a}{1-\gamma a} \right)}
\frac{1}{1 + \gamma x} \sin^2 \left[ n \pi
\frac{\ln \left( \frac{1 + \gamma x}{1 + \gamma a} \right)}
     {\ln \left( \frac{1 + \gamma a}{1 - \gamma a} \right)}
	 \right]
\end{equation}
for $|x| \leq a$, and $\rho_n(x)=0$ otherwise.
As $n\rightarrow \infty$ the classical probability density
$\rho_{{\rm classic}}(x)\rmd x \propto \rmd x/v$
of finding the particle between $x$ and $x+\rmd x$ within well
becomes
\begin{equation}
\label{eq:rho_classic(x)}
\rho_{{\rm classic}}(x) =
\frac{\gamma}{\ln \left(\frac{1+\gamma a}{1-\gamma a}\right)(1+\gamma x)},
\end{equation}
so the uniform probability density $\rho_{{\rm classic}}(x) = 1/(2a)$
is recovered for $\gamma\rightarrow 0$, as expected classically.

The expected values
$ \langle \hat{x} \rangle $,
$ \langle \hat{x}^2 \rangle $,
$ \langle \hat{p} \rangle $ and
$ \langle \hat{p}^2 \rangle $
for the particle in the one-dimensional symmetric infinite
square well are
\numparts
\label{eq:expectation_values_quantum}
\begin{eqnarray}
\label{eq:x-med-quantum}
\langle \hat{x} \rangle
&=& \frac{1}{\gamma}
\left[ \frac{\gamma a}{{\rm atanh}(\gamma a)} - 1 \right]
-\frac{a\,{\rm atanh}(\gamma a)}{{\rm atanh}^2(\gamma a) +
(n{\pi})^2},
\\
\label{eq:x^2-med-quantum}
\langle \hat{x}^2 \rangle
&=& -\frac{1}{\gamma^2}
\left[ \frac{\gamma a}{{\rm atanh}(\gamma a)} - 1 \right]
-\frac{4a\,{\rm atanh}(\gamma a)}{\gamma [4\,{\rm atanh}^2
(\gamma a) + (n\pi)^2 ]}
\nonumber \\
&&
+\frac{2a\,{\rm atanh}(\gamma a)}{\gamma [{\rm atanh}^2(\gamma a)
+ (n\pi)^2 ]},
\\
\label{eq:expected-value-p}
\langle \hat{p} \rangle &=& 0,
\\
\label{eq:expected-value-p^2}
\langle \hat{p}^2 \rangle
&=& \frac{\hbar^2 k_{\gamma,n}^2 \gamma a}{(1-\gamma^2 a^2)^2
\,{\rm atanh}(\gamma a)}
\left[ 1 + \frac{{\rm atanh}^2(\gamma a)}{4\,{\rm atanh}^2
(\gamma a)+(n{\pi})^2}\right].
\end{eqnarray}
\endnumparts
It can be shown that
$\lim_{\gamma\rightarrow 0}\langle \hat{x} \rangle = 0$,
$\lim_{\gamma\rightarrow 0} \langle \hat{x}^2 \rangle
  = \frac{a^2}{3}\left(1-\frac{6}{n^2 \pi^2} \right)$
and
$\lim_{\gamma\rightarrow 0}\langle \hat{p}^2 \rangle
 = \left( \frac{\pi n}{2a} \right)^2$
which correspond to the standard case.
The first moments of the position and the linear
momentum for the classical distribution (\ref{eq:rho_classic(x)}) are
\numparts
\begin{eqnarray}
\label{eq:xmed_classic}
	\overline{x} &=&
     \frac{1}{\gamma }
	 \left[ \frac{\gamma a}{{\rm atanh}(\gamma a)} - 1 \right],
\\
\label{eq:xquadmed_classic}
	\overline{x^2}
      	&=& -\frac{1}{\gamma^2}
	      \left[ \frac{\gamma a}{{\rm atanh}(\gamma a)} - 1 \right],
\\
\label{eq:pmed_classic}
\overline{p}  &=& 0,
\\
\label{eq:pquadmed_classic}
\overline{p^2} &=& \frac{2m_0 E \gamma a}{(1-\gamma^2 a^2)^2\,
				   {\rm atanh} (\gamma a)},
\end{eqnarray}
\endnumparts
where $\lim_{\gamma\rightarrow 0} \overline{x}=0$,
$\lim_{\gamma\rightarrow 0} \overline{x^2}=a^2/3$ and
$\lim_{\gamma\rightarrow 0} \overline{p^2}=2m_0 E$.
It is straightforward to verify that
(\ref{eq:x-med-quantum}) to (\ref{eq:expected-value-p^2})
coincide  respectively with
(\ref{eq:xmed_classic}) to (\ref{eq:pquadmed_classic}) as $n\rightarrow \infty$, thus
showing the classical limit in terms of probability densities.

\begin{figure}[!h]
\centering
\begin{minipage}[b]{0.32\linewidth}
\includegraphics[width=\linewidth]{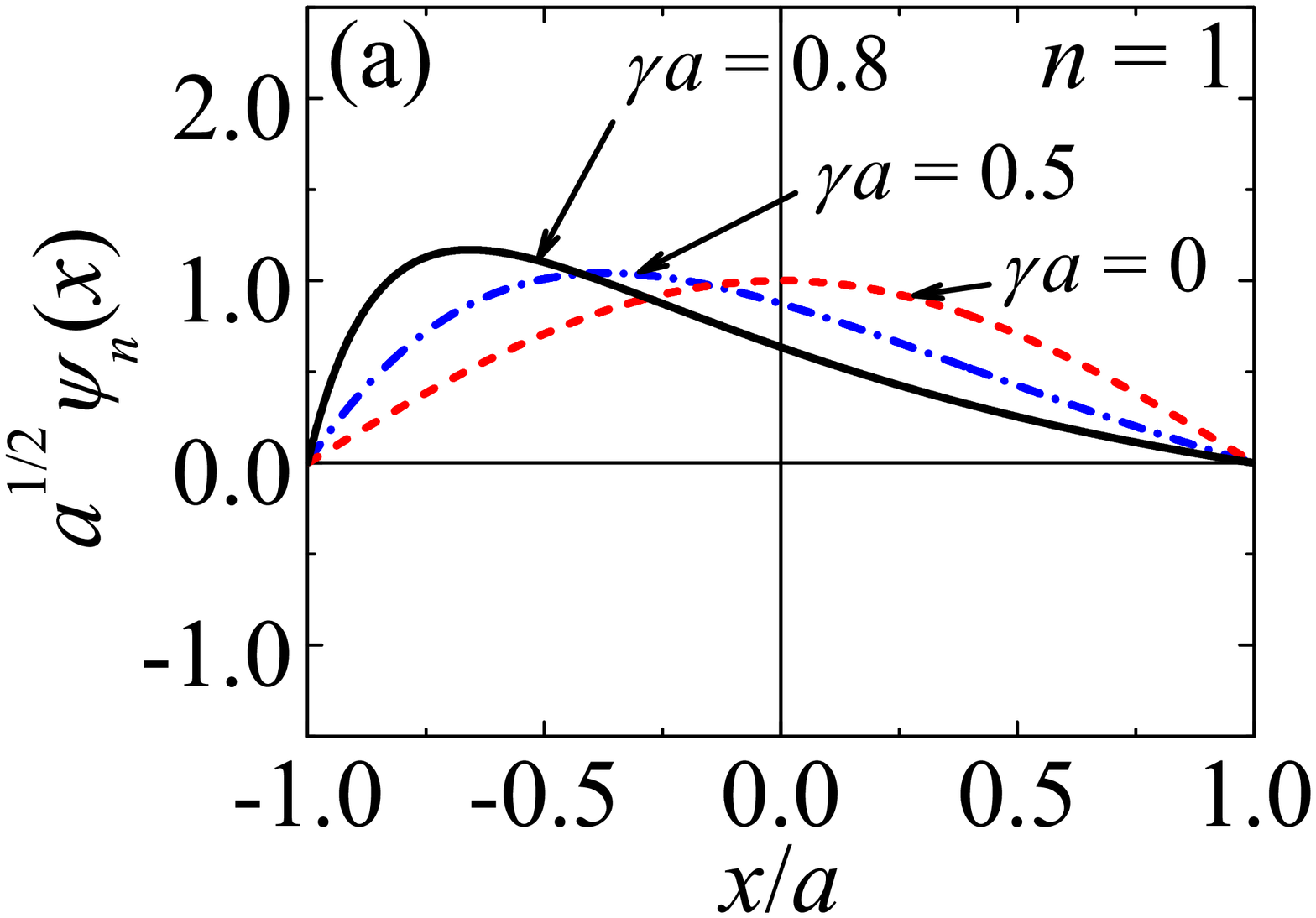}
\end{minipage}
\begin{minipage}[b]{0.32\linewidth}
\includegraphics[width=\linewidth]{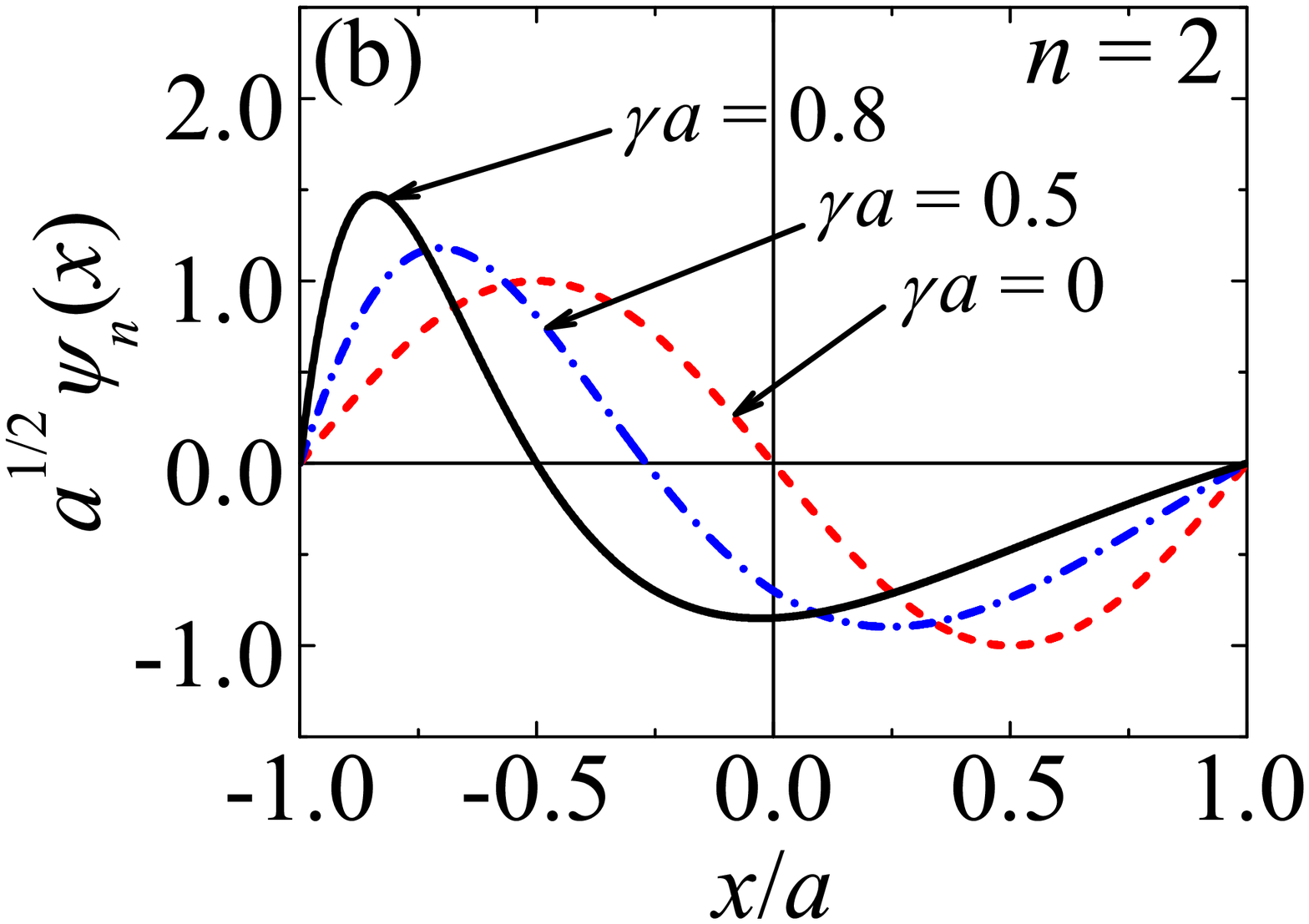}
\end{minipage}
\begin{minipage}[b]{0.32\linewidth}
\includegraphics[width=\linewidth]{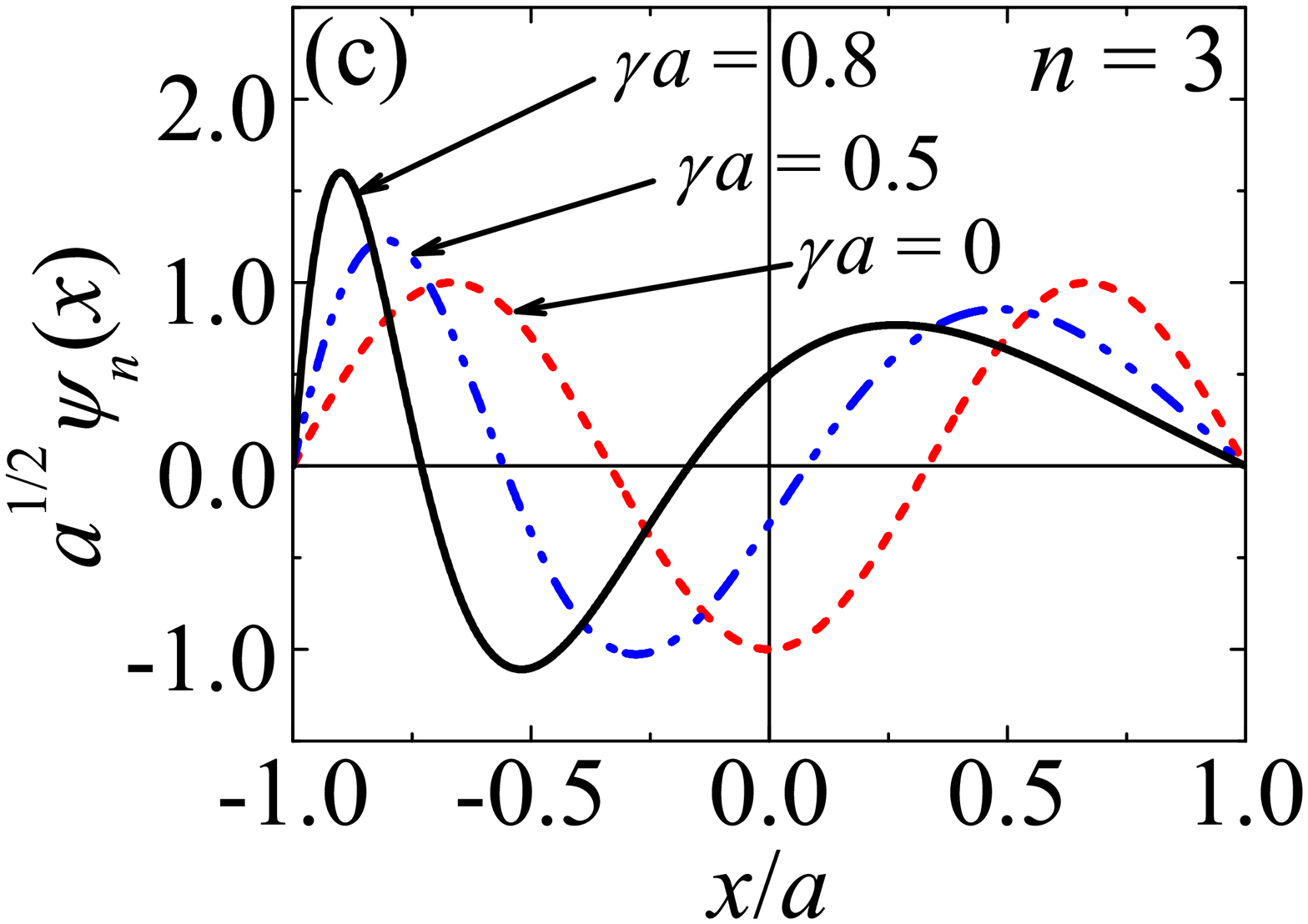}
\end{minipage} \\
\begin{minipage}[b]{0.32\linewidth}
\includegraphics[width=\linewidth]{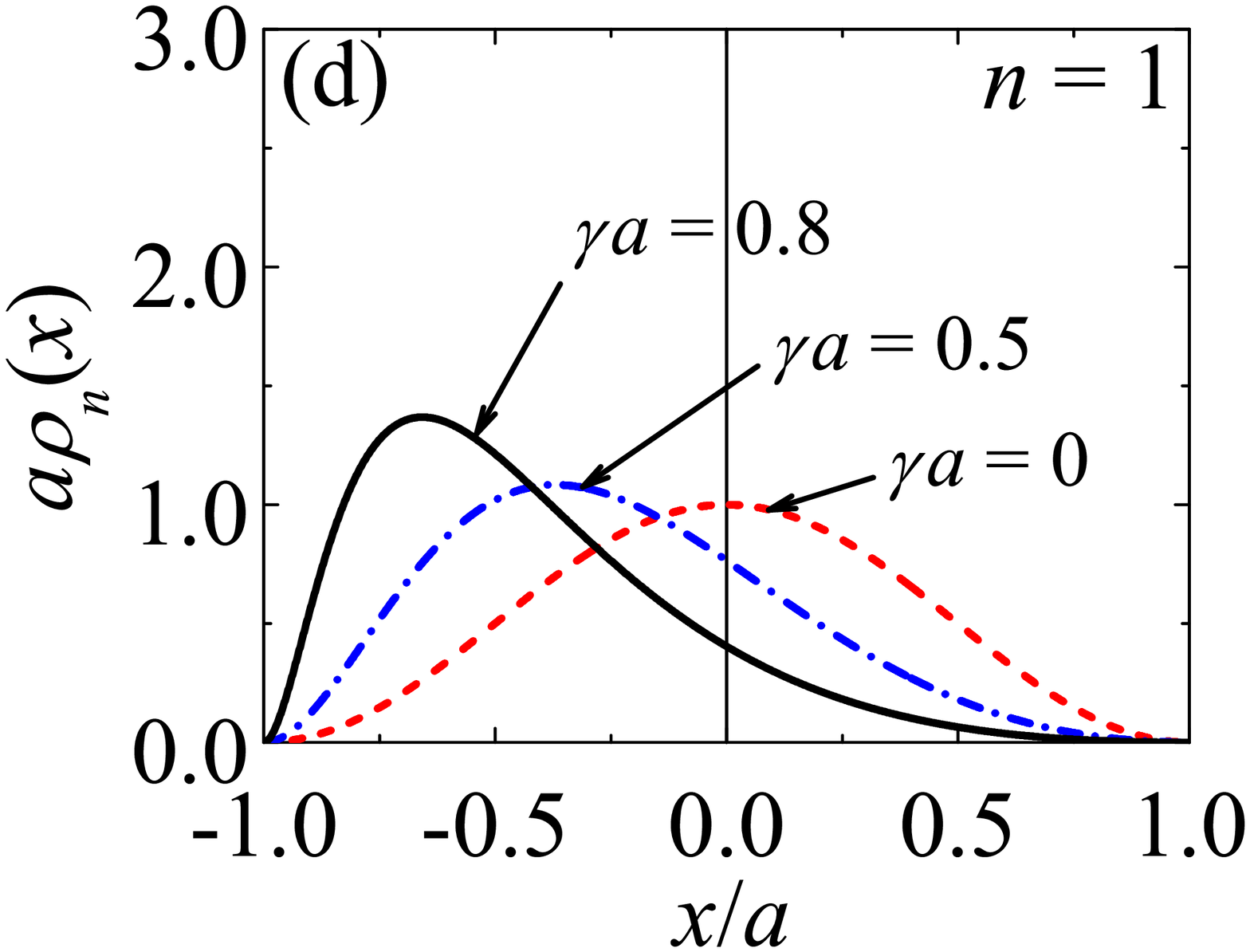}
\end{minipage}
\begin{minipage}[b]{0.32\linewidth}
\includegraphics[width=\linewidth]{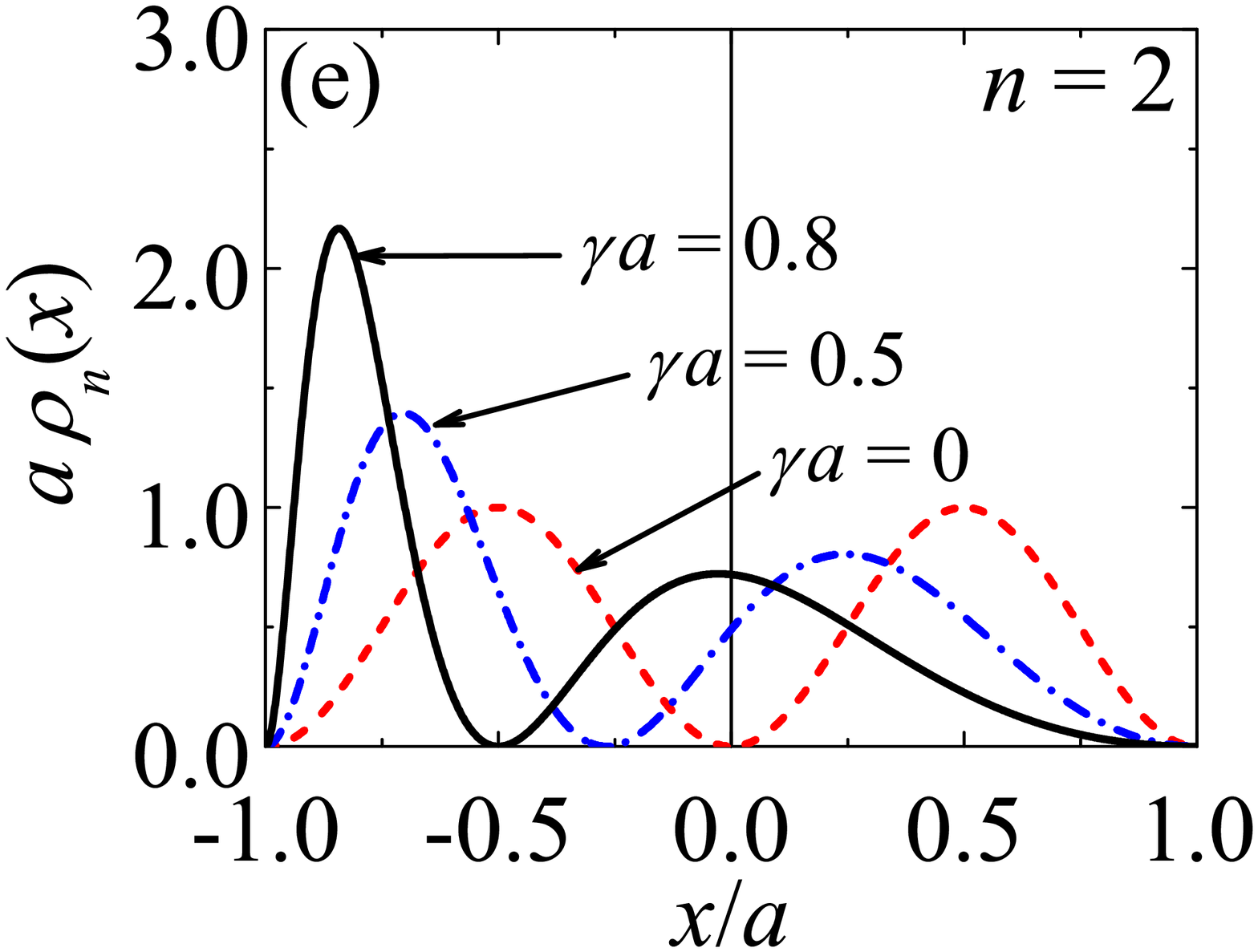}
\end{minipage}
\begin{minipage}[b]{0.32\linewidth}
\includegraphics[width=\linewidth]{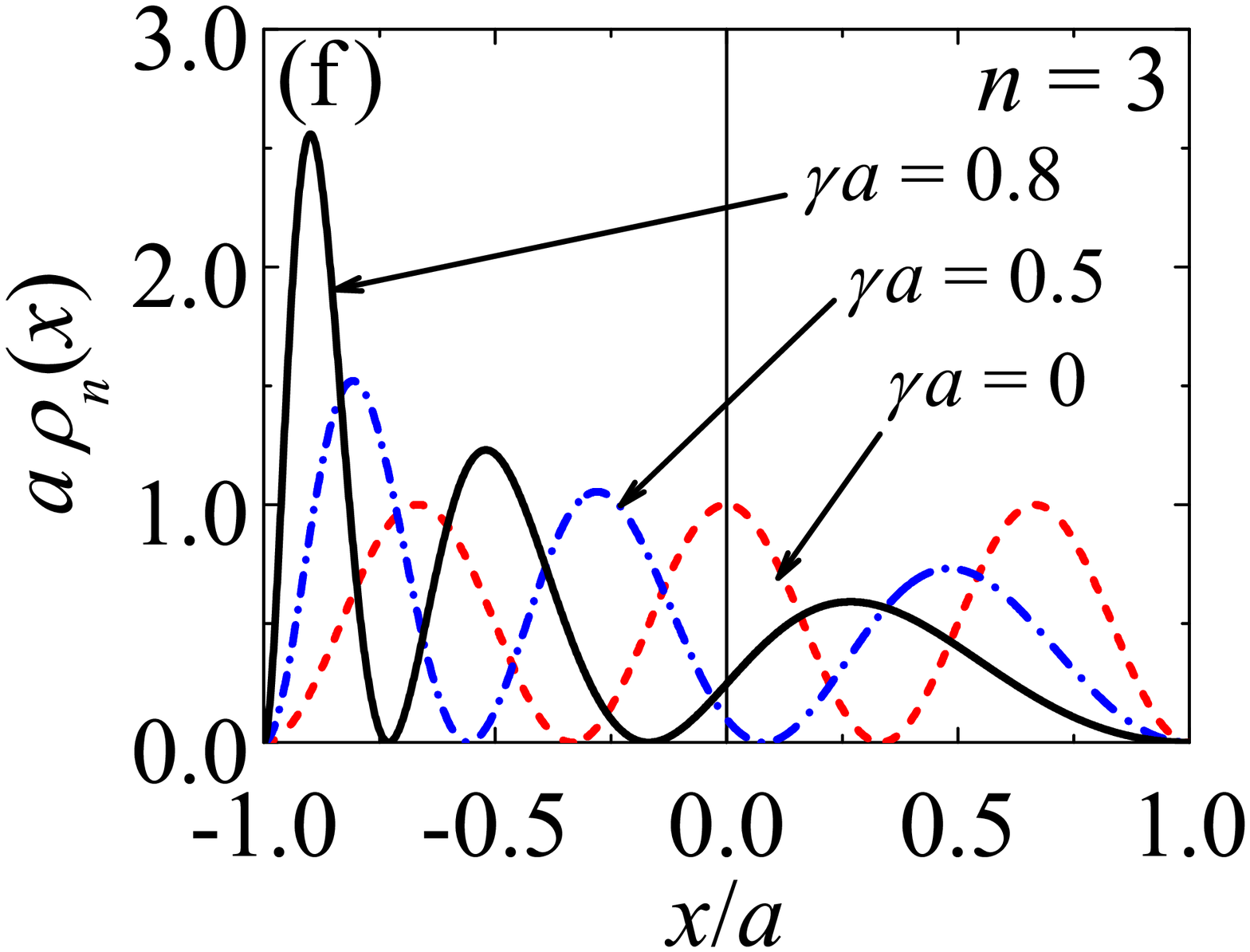}
\end{minipage} \\
\begin{minipage}[b]{0.32\linewidth}
\includegraphics[width=\linewidth]{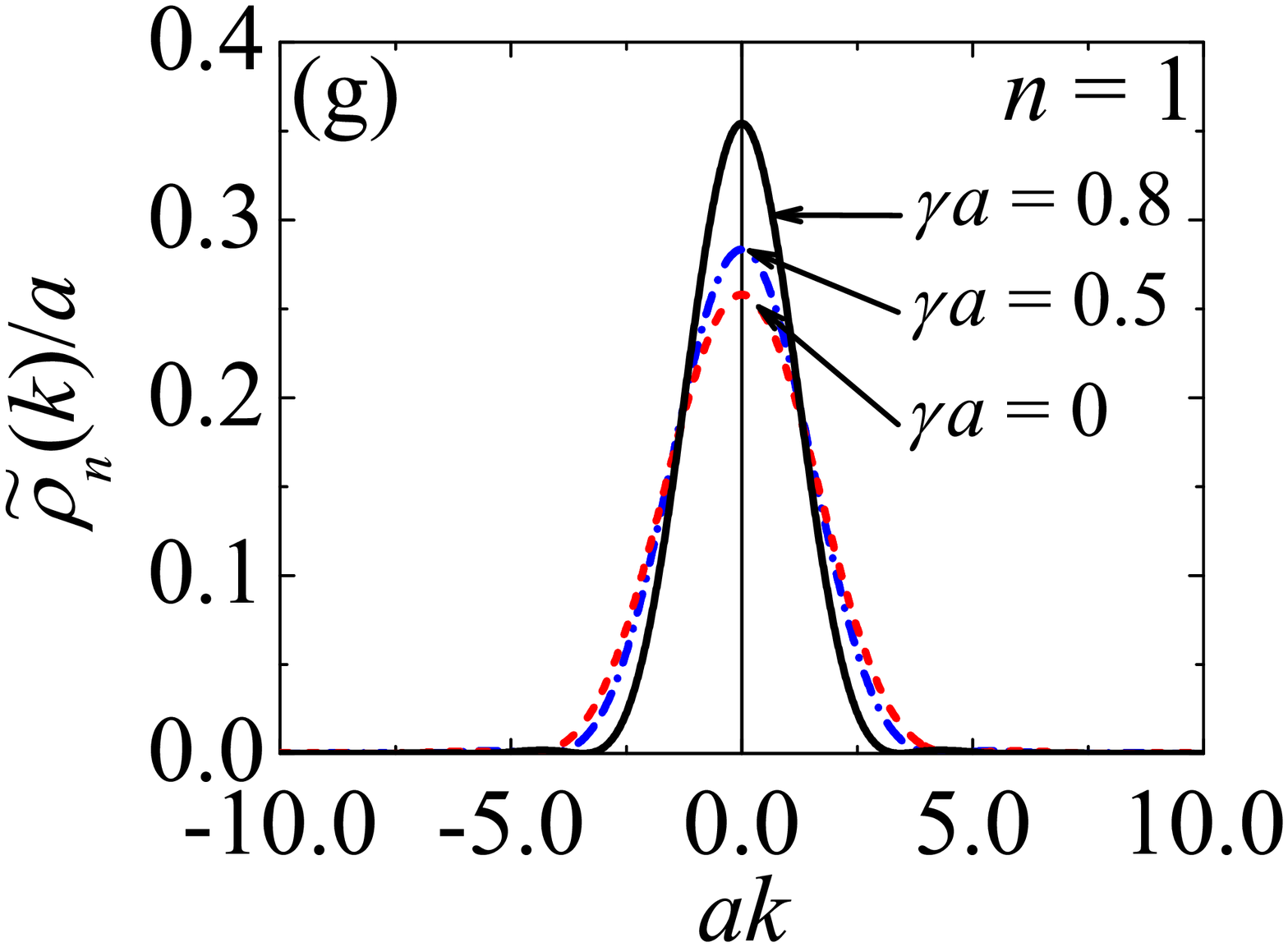}
\end{minipage}
\begin{minipage}[b]{0.32\linewidth}
\includegraphics[width=\linewidth]{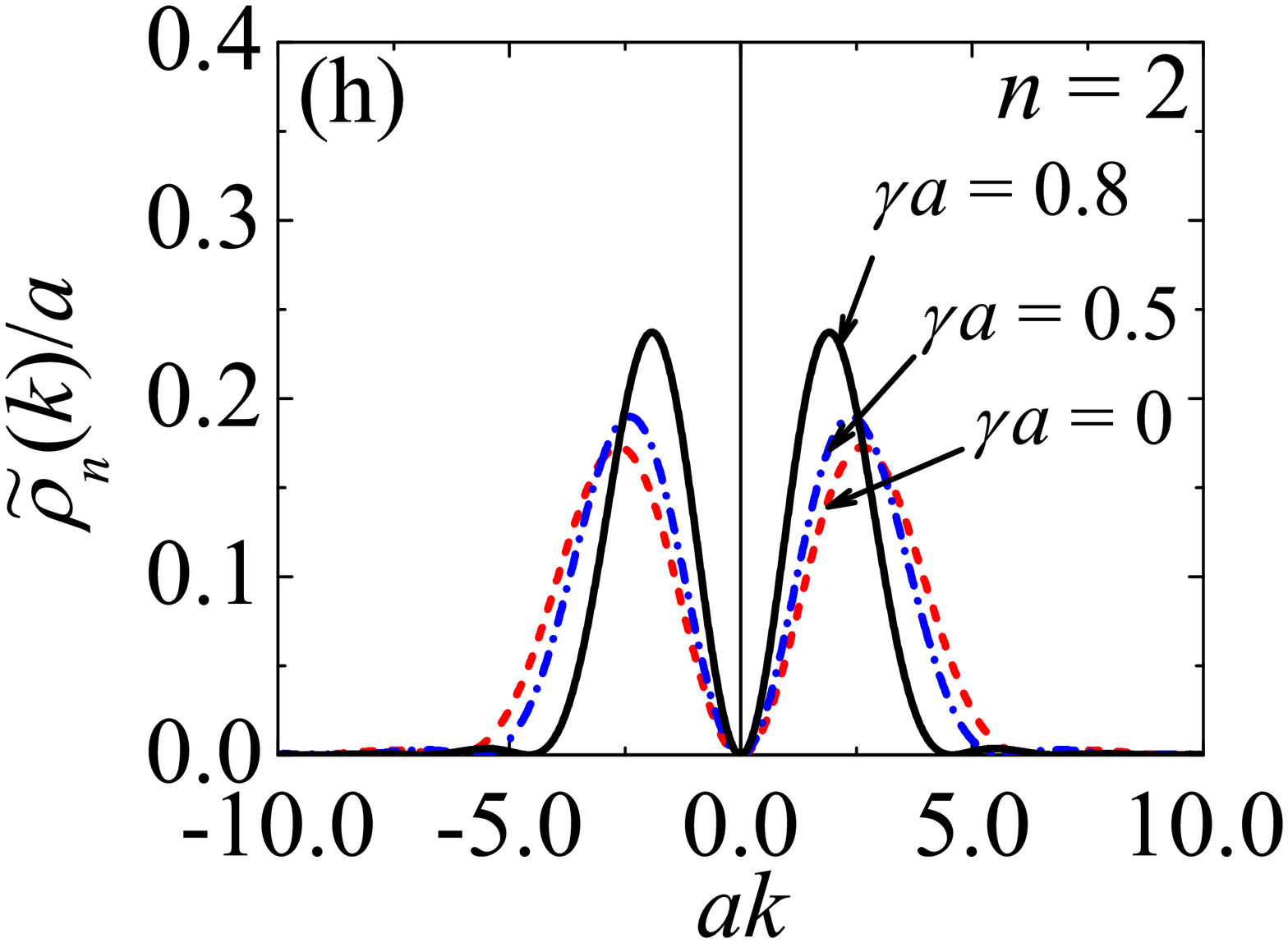}
\end{minipage}
\begin{minipage}[b]{0.32\linewidth}
\includegraphics[width=\linewidth]{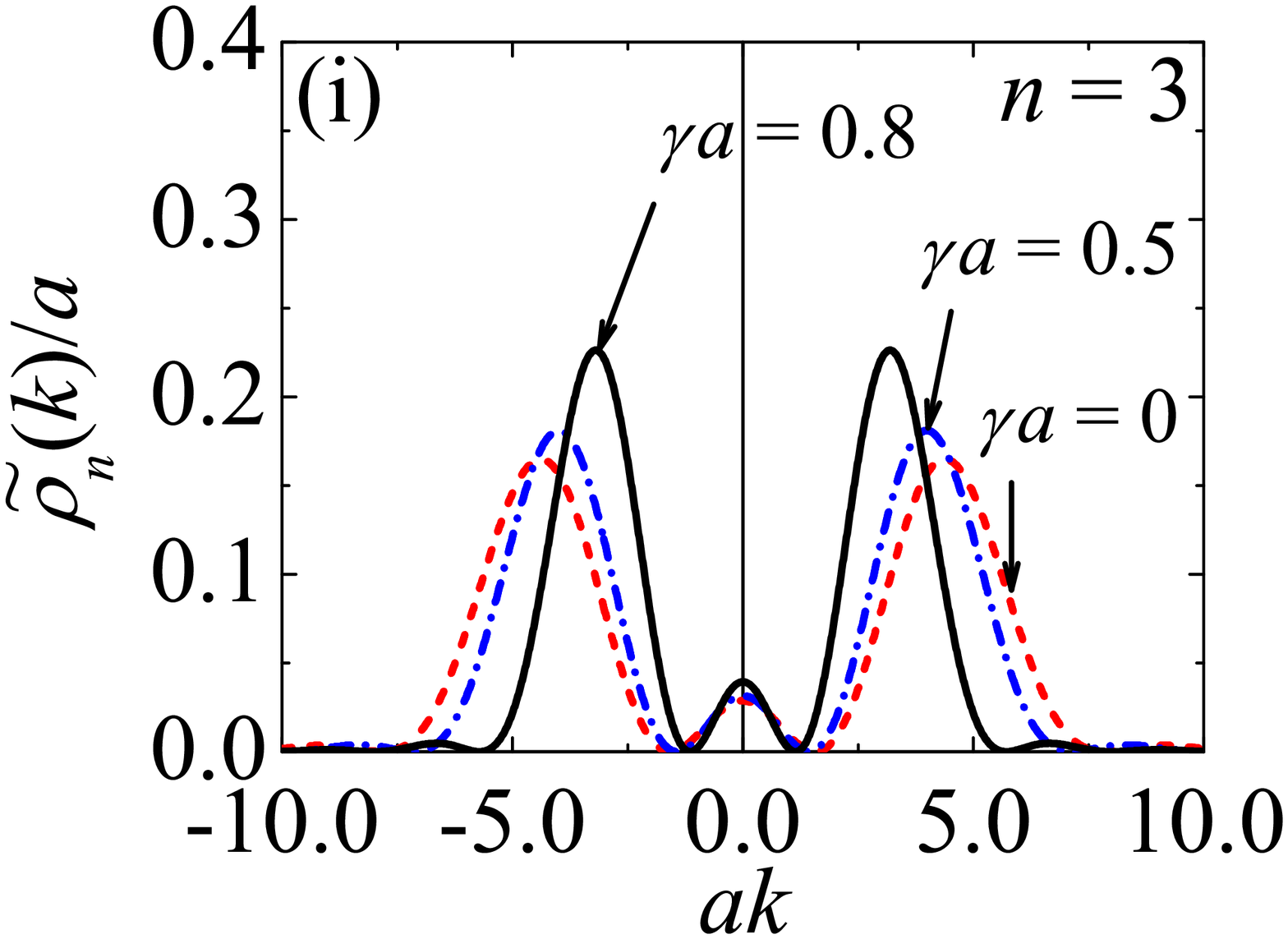}
\end{minipage} \\
\caption{\label{fig:1}
(Color online) The first three energy eigenfunctions $\psi_n(x)$
(first line)
along with their probability densities $\rho_n(x)$ (second line)
and $\widetilde{\rho}_n(k)$ (third line) for a particle with a PDM
given by (\ref{eq:m(x)}) in a symmetric infinite square well, and
for different values of $\gamma a$
(for comparing, the standard case $\gamma a = 0$ is also
represented).
(a), (d) and (g): $n = 1$ (ground state).
(b), (e) and (h): $n = 2$ (first excited state).
(c), (f) and (i): $n = 3$ (second excited state).
}
\end{figure}

\begin{figure}[!h]
\centering
\begin{minipage}[b]{0.40\linewidth}
\includegraphics[width=\linewidth]{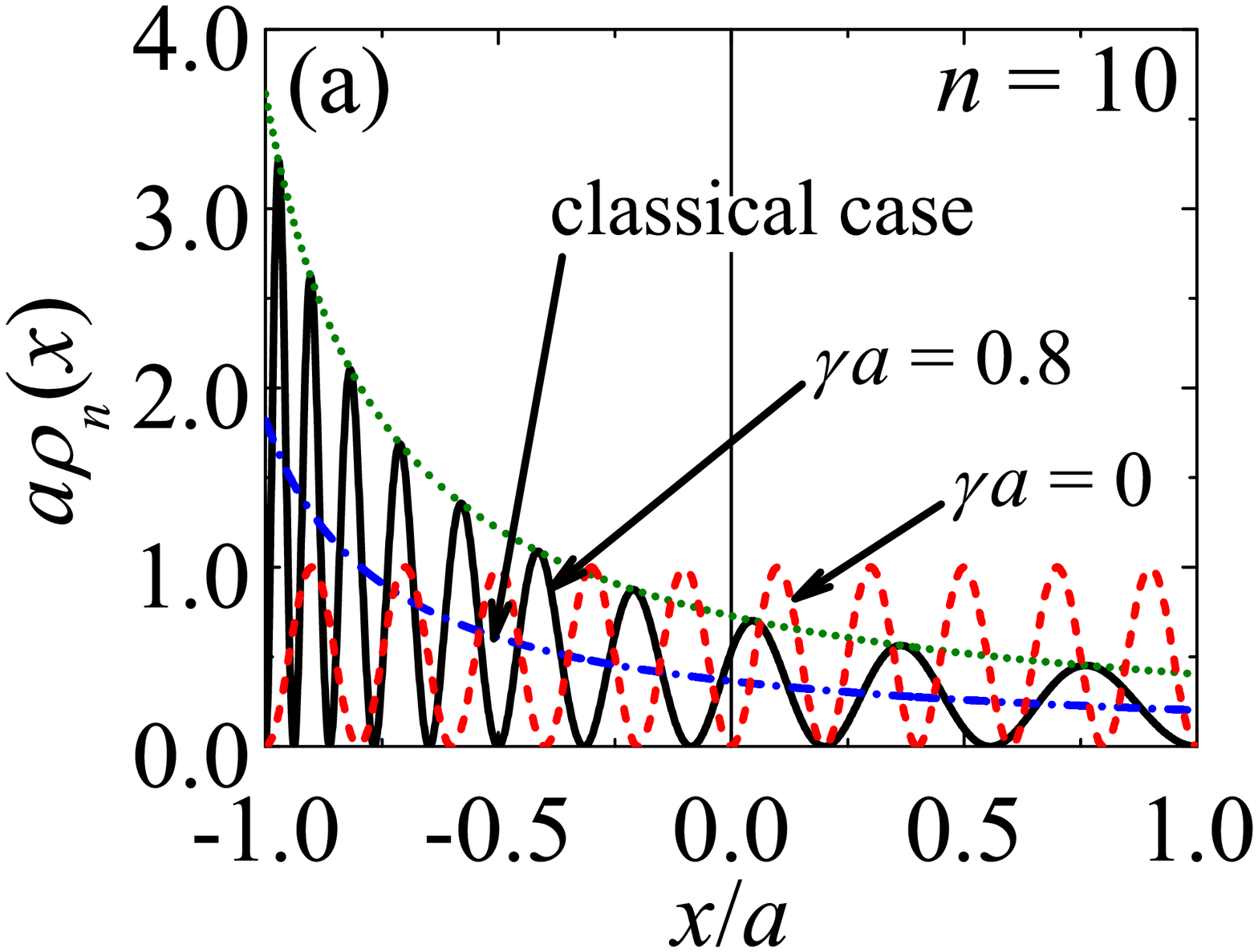}
\end{minipage}
\begin{minipage}[b]{0.40\linewidth}
\includegraphics[width=\linewidth]{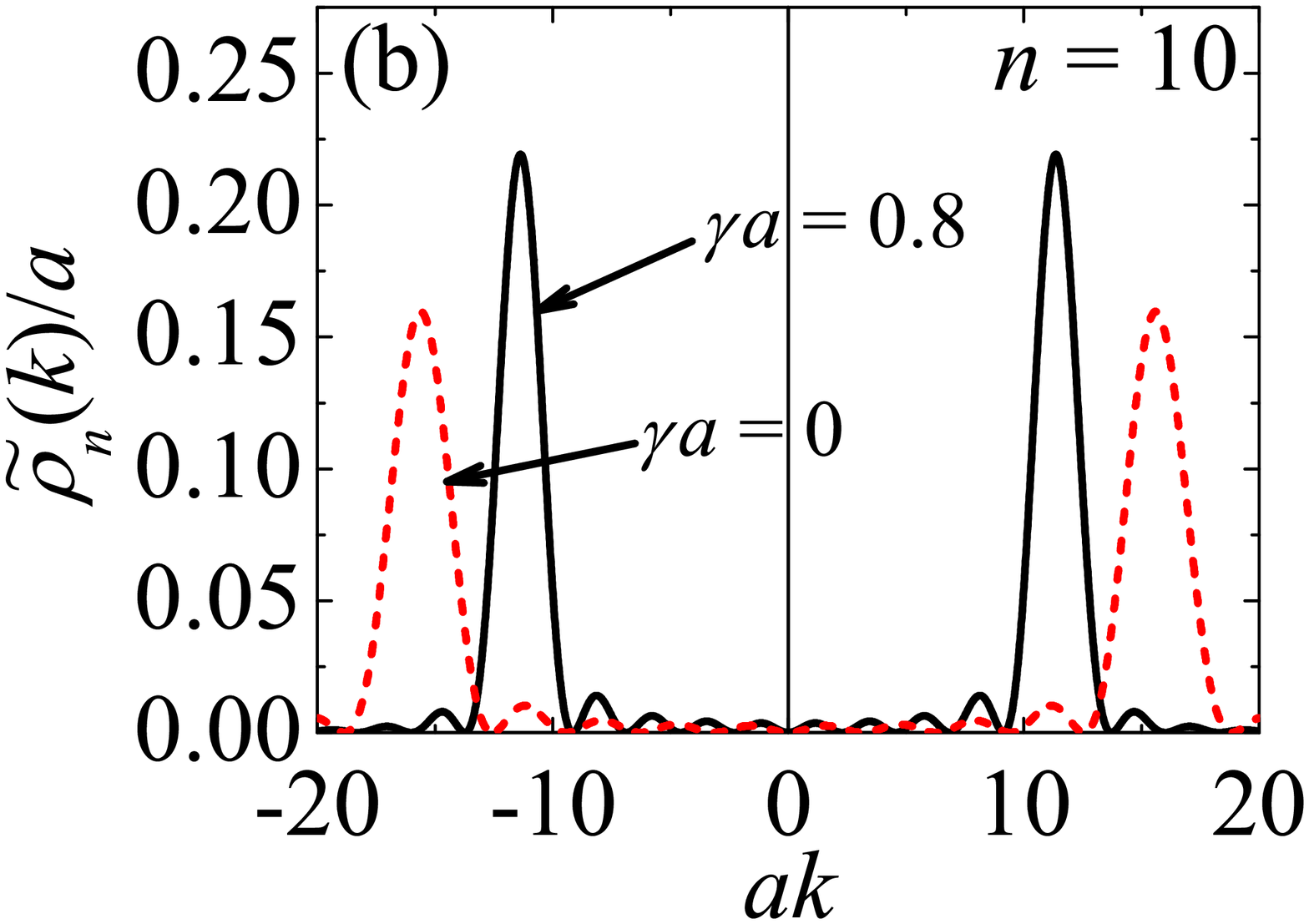}
\end{minipage}
\caption{\label{fig:2}
(Color online) Probability densities (a) $\rho_n(x)$ and (b)
$\widetilde{\rho}_n(k)$ for a PDM (\ref{eq:m(x)}) in a symmetric
infinite square well for $\gamma a = 0.8$. and $n=10$. The
classical distribution $\rho_{{\rm classical}}(x)$
(dot-dashed curve) given by (\ref{eq:rho_classic(x)})
is shown for comparison in (a).
The upper bound (dotted curve) corresponds to
$2\rho_{{\rm classical}}(x)$. In both figures (a) and (b), the
usual case ($\gamma a = 0$) is shown for comparison.}
\end{figure}

Substituting the eigenfunctions (\ref{eq:psi_n}) in
(\ref{eq:inverse-deformed-transform}), we obtain
the eigenfunctions in $k$-space
\begin{equation}
\label{eq:psi_k-space}
\widetilde{\psi}_n (k) =
	\sqrt{\frac{\pi n^2 L_\gamma}{4}}
    \frac{\sin \left( \frac{k L_\gamma}{2} -\frac{n\pi}{2} \right)
    }{
	\left( \frac{kL_\gamma}{2} \right)^2
	- \left( \frac{n \pi}{2}\right)^2}
	\rme^{-\frac{\rmi}{2} [k\gamma^{-1} \ln (1-\gamma^2 a^2)
    + \pi (n+1)]},
\end{equation}
and consequently, the corresponding probability densities are
\begin{equation}
\label{eq:rho_n(k)}
\widetilde{\rho}_n (k) = |\widetilde{\psi}_n (k)|^2 =
	\frac{\pi n^2 L_\gamma}{4}
    \frac{\sin^2 \left( \frac{k L_\gamma}{2} -\frac{n\pi}{2}
    \right)}{
	\left[\left( \frac{kL_\gamma}{2} \right)^2
	- \left( \frac{n \pi}{2}\right)^2 \right]^2}.
\end{equation}
From the probability densities (\ref{eq:rho_n(k)}),
we have
\numparts
\begin{eqnarray}
\label{eq:expected-value-k}
\langle \hat{\Pi} \rangle &=& \hbar \langle \hat{k} \rangle = 0,
\\
\label{eq:expected-value-k^2}
\langle \hat{\Pi}^2 \rangle &=&
\hbar^2\langle \hat{k}^2 \rangle =
\left( \frac{n \pi \hbar}{L_\gamma} \right)^2.
\end{eqnarray}
\endnumparts
\noindent Figure \ref{fig:1} illustrates the energy eigenstates
for the three states of the lowest energies with different values
of $\gamma a$, as well as their probability densities in the $x$
and $k$ spaces.
For $n=10$, we can see from the figure \ref{fig:2} that
the average value of the quantum probability density
(\ref{eq:rho_n(x)}) approaches to the classical probability
density (\ref{eq:rho_classic(x)}), in according to the
correspondence principle.
We also see that when $\gamma a$ increases the mass variation
(in relation with the position) grows, which implies
that the moment distribution turns out more localised around $k=0$.


\subsection{Complexities analysis and uncertainty relations}

\begin{figure*}[!htb]
\centering
\begin{minipage}[b]{0.32\linewidth}
\includegraphics[width=\linewidth]{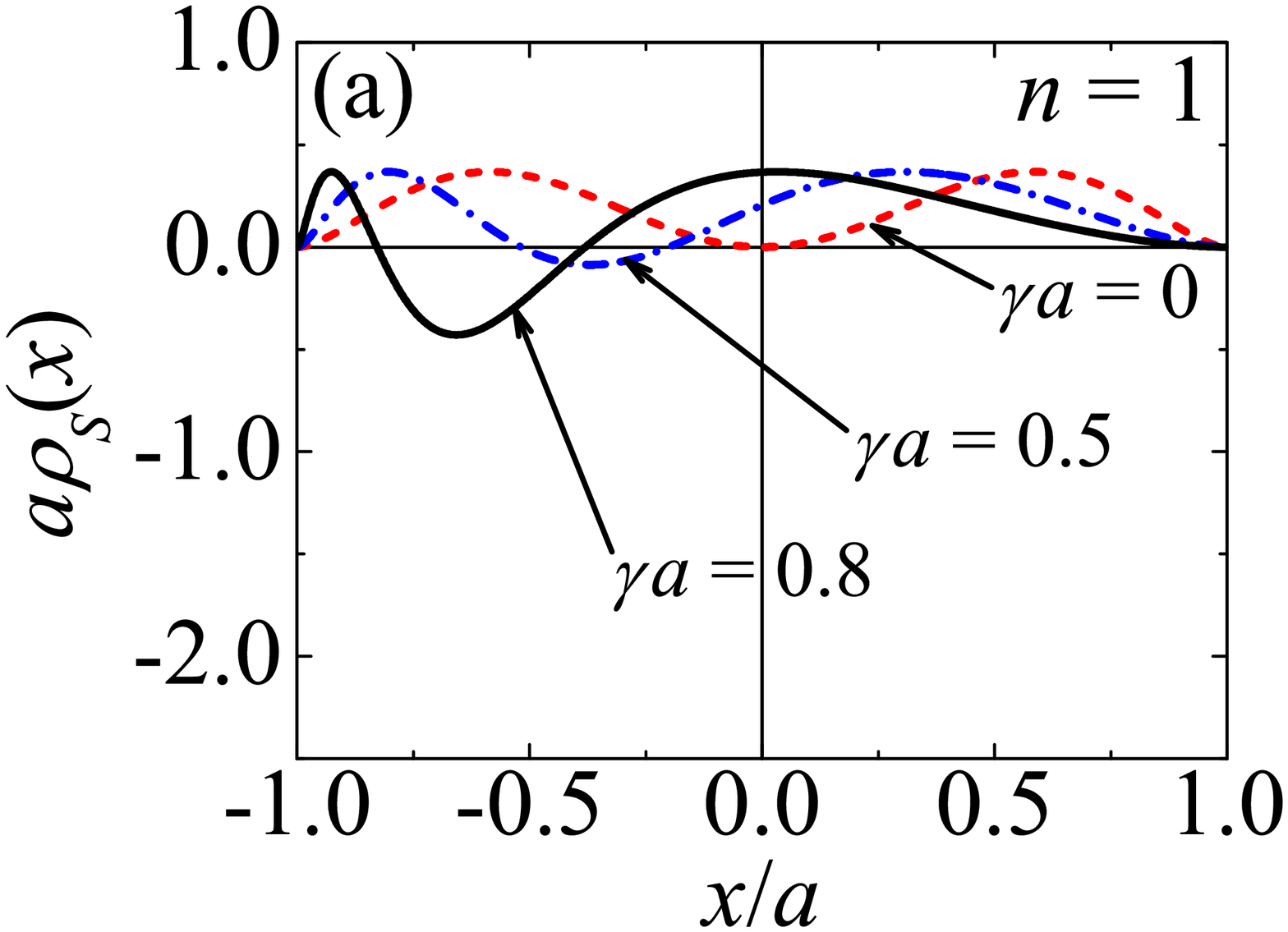}
\end{minipage}
\begin{minipage}[b]{0.32\linewidth}
\includegraphics[width=\linewidth]{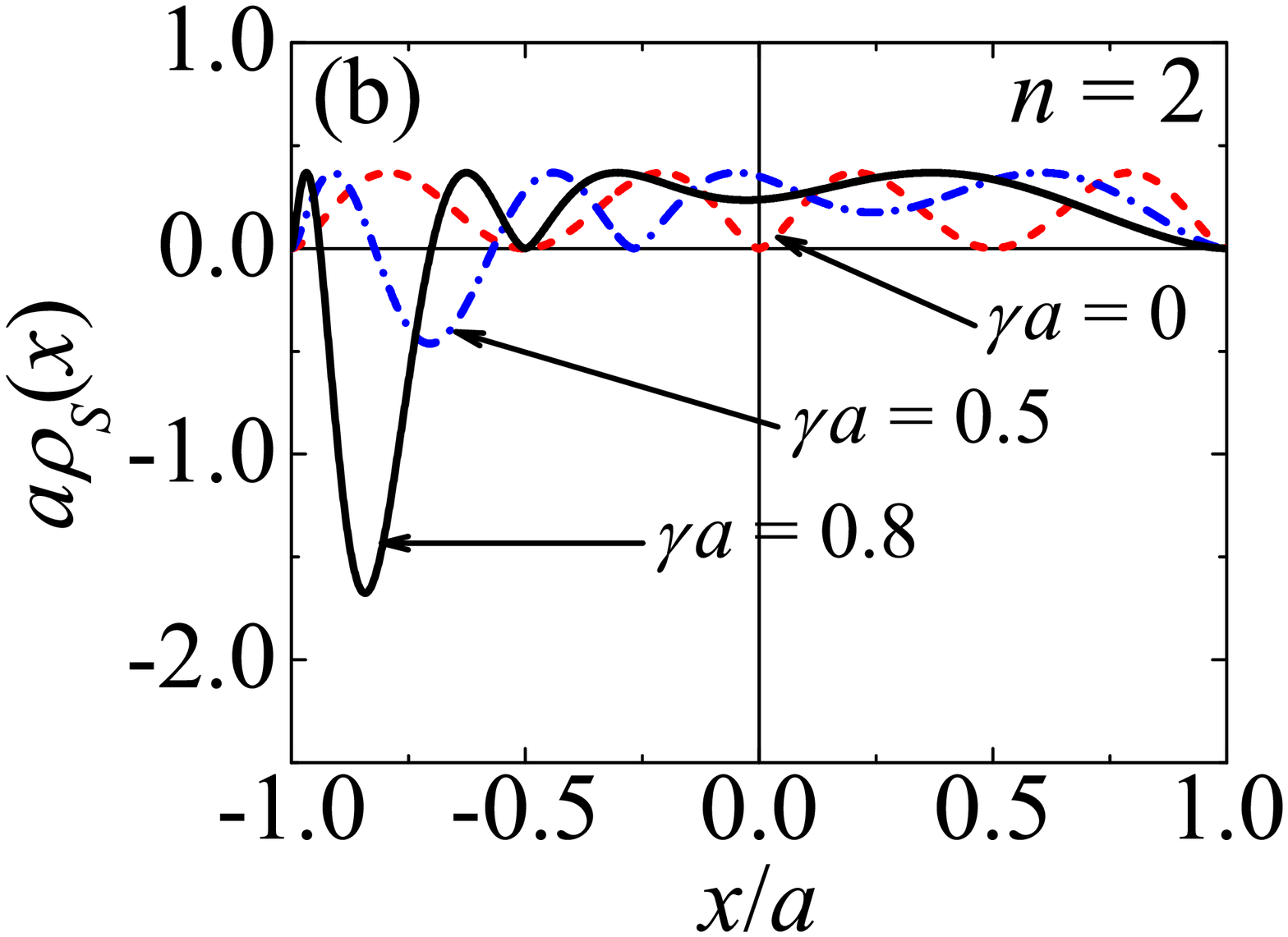}
\end{minipage}
\begin{minipage}[b]{0.32\linewidth}
\includegraphics[width=\linewidth]{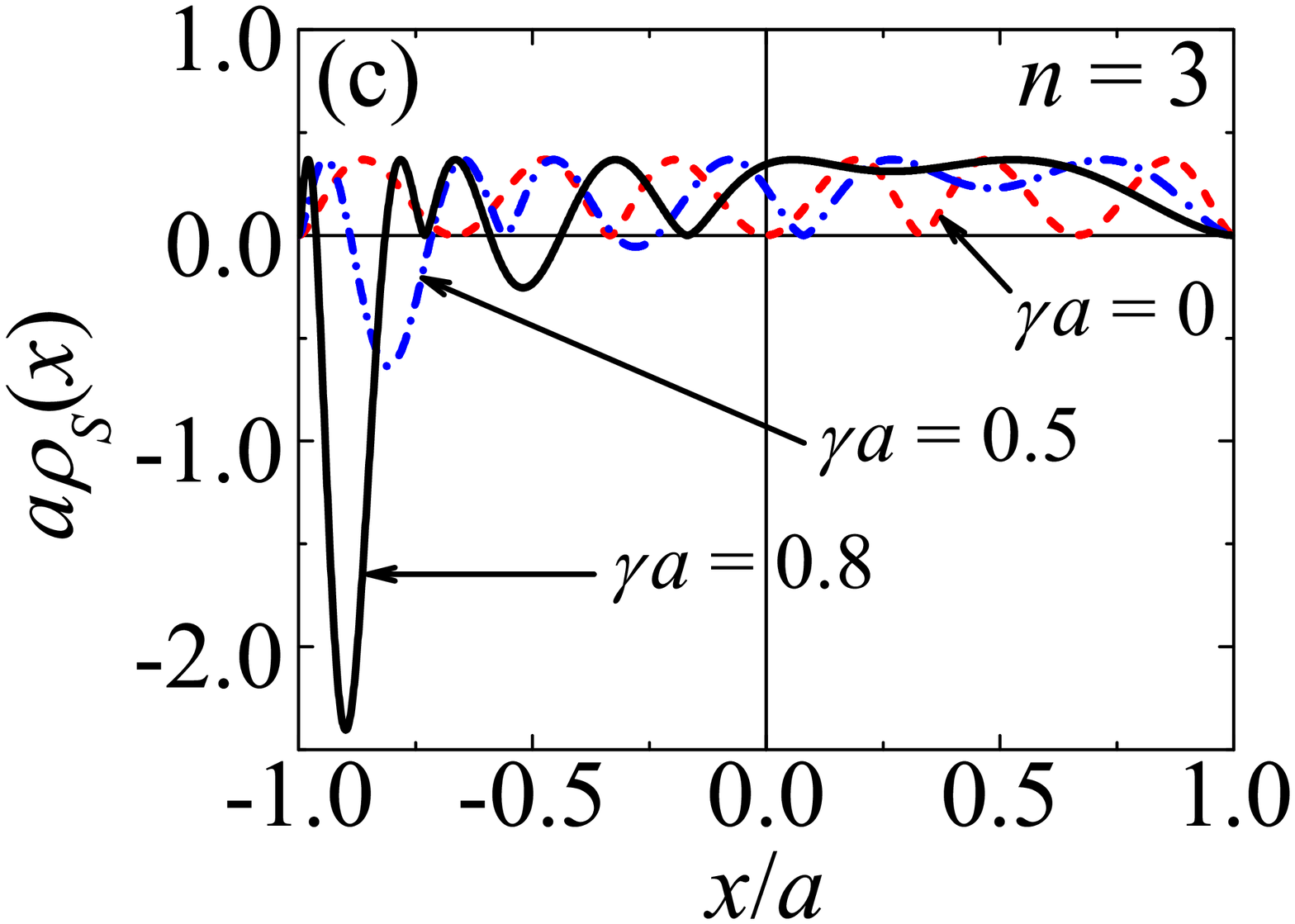}
\end{minipage}\\
\begin{minipage}[b]{0.32\linewidth}
\includegraphics[width=\linewidth]{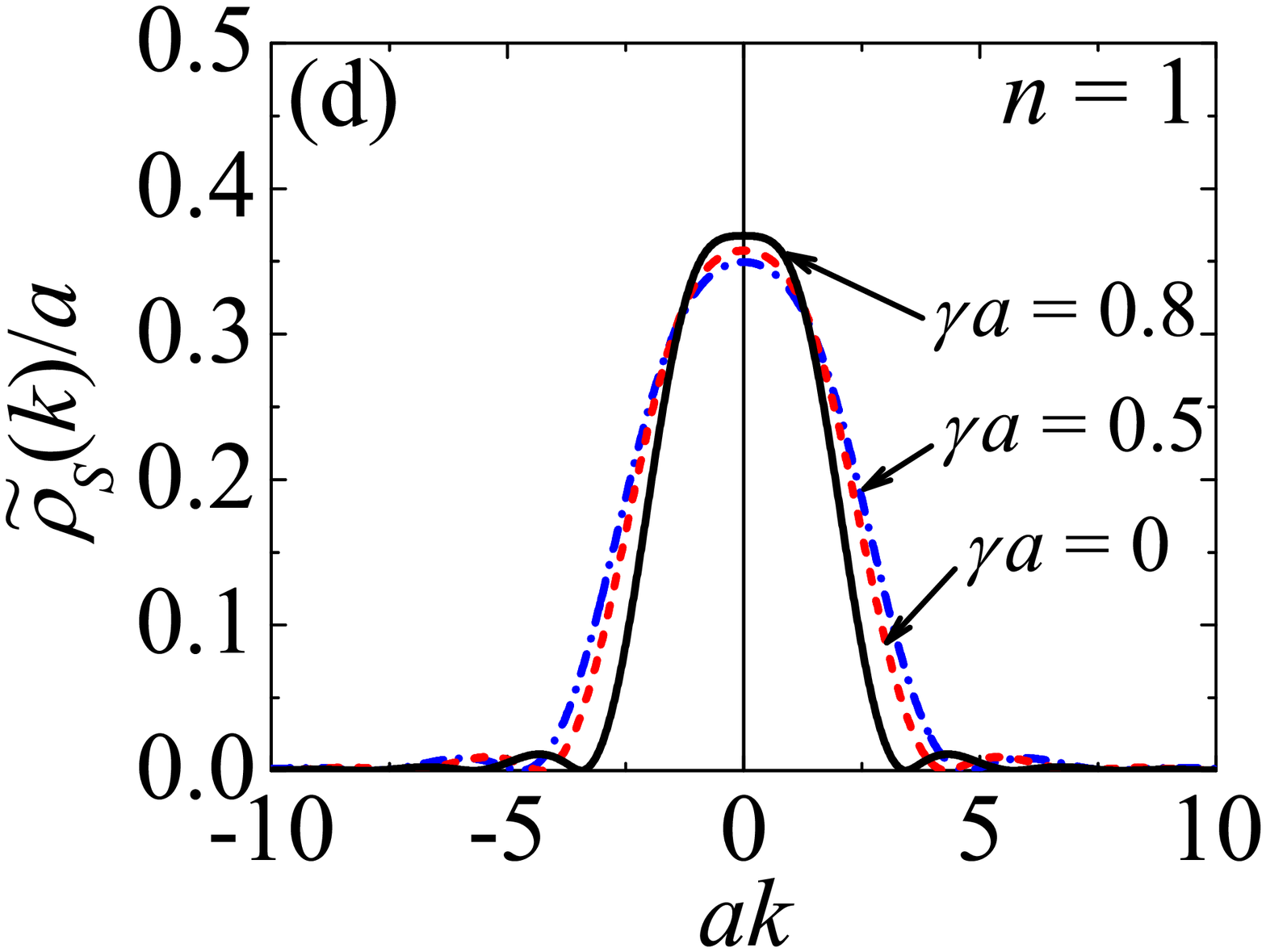}
\end{minipage}
\begin{minipage}[b]{0.32\linewidth}
\includegraphics[width=\linewidth]{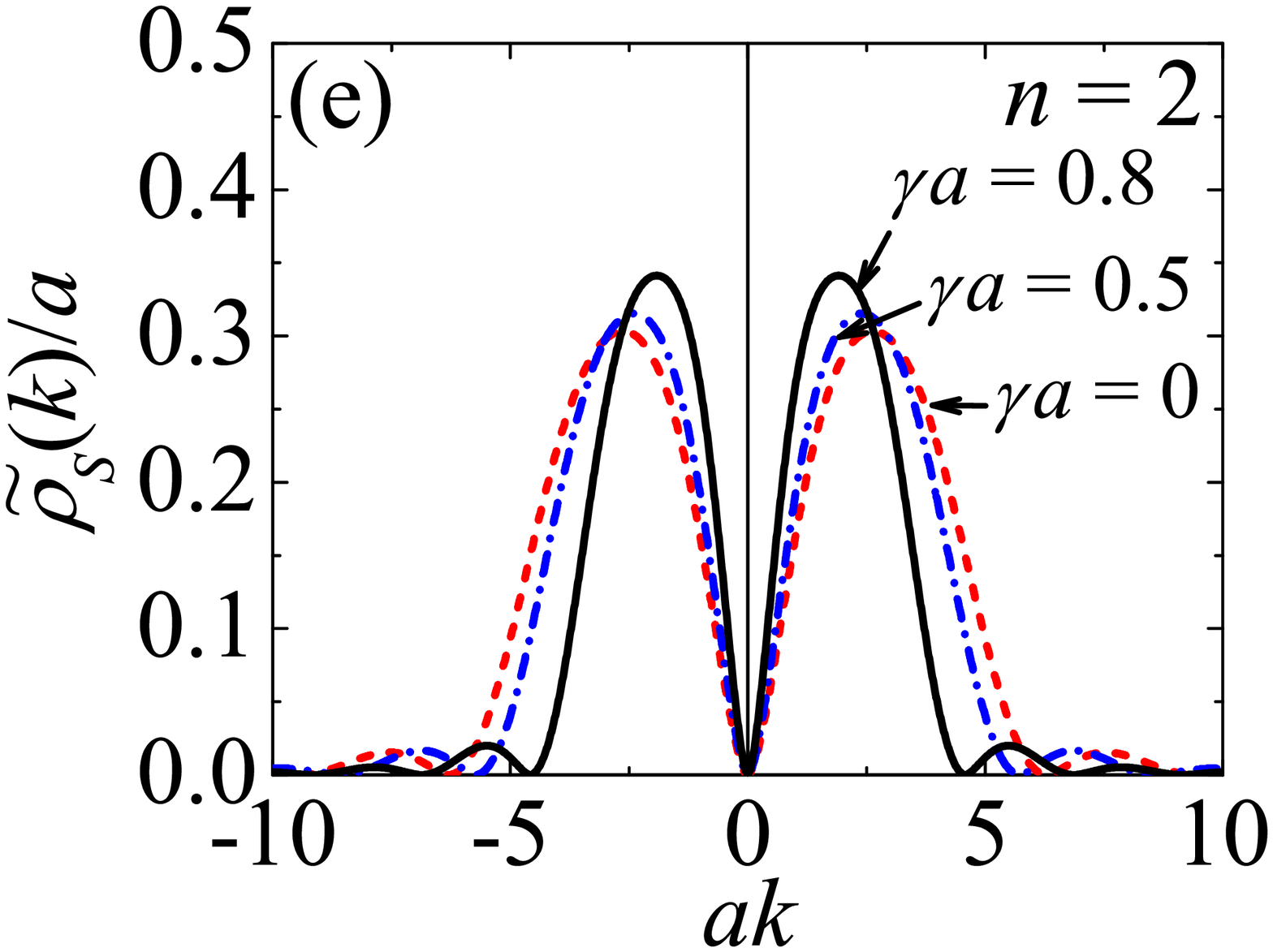}
\end{minipage}
\begin{minipage}[b]{0.32\linewidth}
\includegraphics[width=\linewidth]{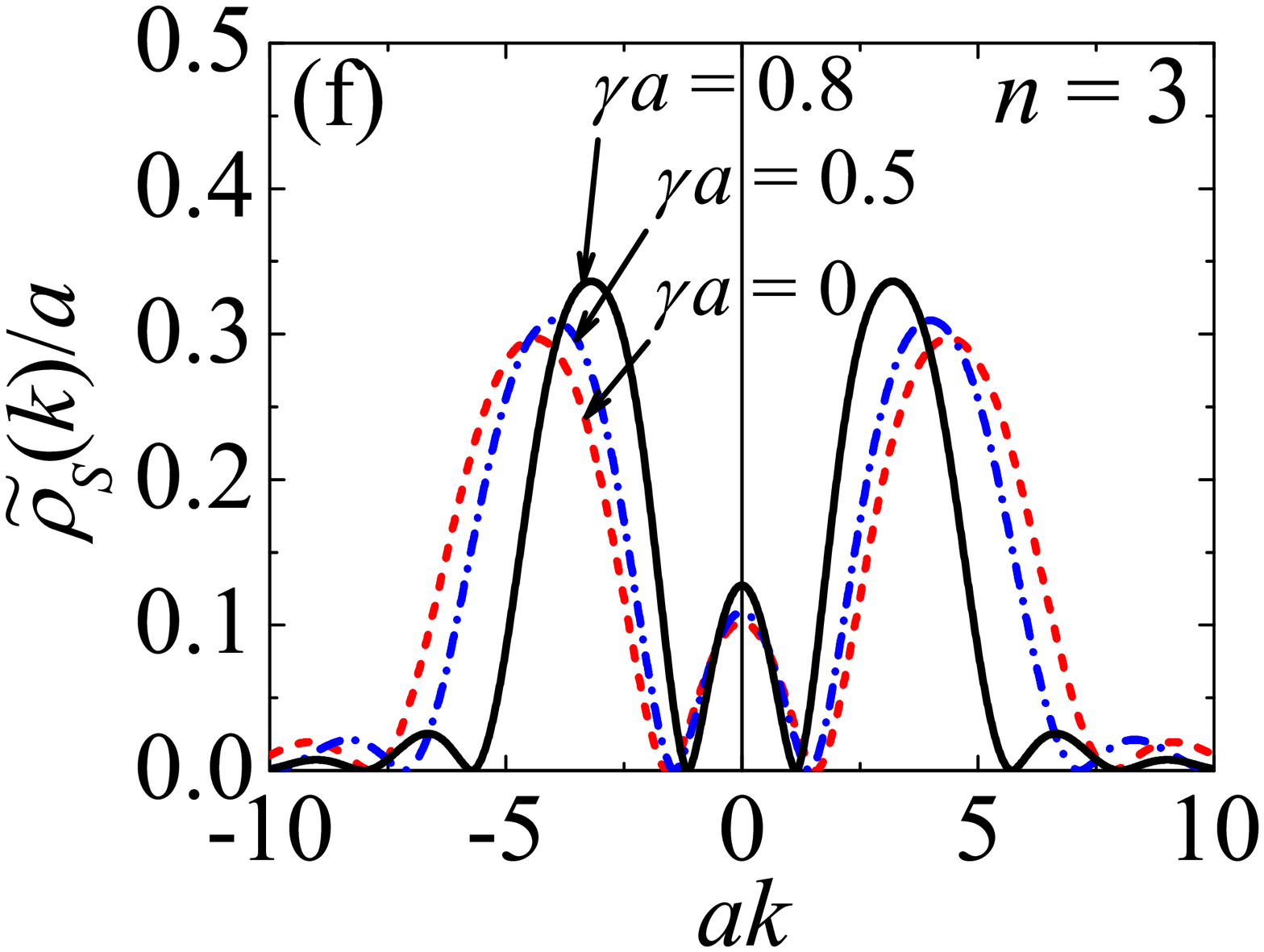}
\end{minipage}
\caption{\label{fig:3}
(Color online) Entropy densities in position (upper line) and
wave-vector (bottom line) spaces for a particle with a PDM
in an infinite potential quantum well for different parameters
$\gamma a$ (the usual case, $\gamma a = 0$, is shown for comparison).
[(a) and (d)] $n = 1$ (ground state),
[(b) and (e)] $n = 2$ (first excited state),
[(e) and (f)] $n = 3$ (second excited state).
As a result of the deformation $\gamma a$ and due to the functional
form of the PDM, the entropy densities
present regions with negative values, which become more pronounced as
the deformation and the quantum number increase.}
\end{figure*}

Inserting the eigenfunctions (\ref{eq:rho_n(x)}) and
(\ref{eq:rho_n(k)}) in (\ref{eq:S_x}), (\ref{eq:S_k}), and
considering $\sigma=a$, we obtain the quantum entropy densities
in the position and the wave-vector spaces
$\rho_{S}(x) \equiv -\rho_n(x) \ln [a \rho_n(x)]$ and
$\rho_{S}(k) \equiv -\widetilde{\rho}_n(k)
\ln [\widetilde{\rho}_n(k)/a]$.
Figure \ref{fig:3} shows the quantum entropy densities, in the
position and the wave-vector representations, for the three first
eigenstates and some values of the deformation parameter
$\gamma a$. The entropy in position space is
\begin{equation}
\label{eq:S_x-infinite-well}
\mathcal{S}_x
[{\rho}_n] = \ln \left( \frac{2 L_{\gamma}}{a}
                        \sqrt{1 - \gamma^2 a^2} \right) - 1
\end{equation}
from which the usual case
$\mathcal{S}_x[\rho_n] = \ln (4) - 1$
is recovered for $\gamma \rightarrow 0$.
For the case of particles with constant mass,
the independence of the entropy
$\mathcal{S}_x[\rho_n]$
with $n$ is due to the periodicity of the probability density
$\rho_n(x)$ (see Ref. \cite{Lopez-Rosa-2011}).
Note that although the probability density
is non-periodic the entropy in position
(\ref{eq:S_x-infinite-well}) does not depend on $n$.
From the classical probability distribution
(\ref{eq:rho_classic(x)}) we obtain
$\mathcal{S}_x[{\rho}_{{\rm classical}}]
= \ln (L_\gamma \sqrt{1-\gamma^2 a^2}/a)$,
and then it satisfies the relation between the entropies for
the distributions in the classical and quantum formalisms, i.e.,
$\mathcal{S}_x [{\rho}_n] = \mathcal{S}_x [{\rho}_{{\rm classical}}]
+ \ln (2) -1$
\cite{Sanchez-Ruiz-1997}.

For the entropy in the $k$-space, using the probability density
(\ref{eq:rho_n(k)}) we have
\begin{equation}
\label{eq:S_k-infinite-well}
\mathcal{S}_k
[\widetilde{\rho}_n] =
			- \ln \left( \frac{2 L_\gamma}{a} \right)
			+ f(n),
\end{equation}
which $f(n)$ is a trigonometric functional given by
\begin{equation}
	f(n) = \ln \left( \frac{8}{\pi} \right)
           - \pi \int_{-\frac{\pi n}{2}}^{+\infty}
			\frac{n^2 \sin^2 u}{(u^2 + \pi n u)^2}
			\ln \left[\frac{n^2 \sin^2 u}{(u^2 + \pi n u)^2}\right] \rmd u.
\end{equation}
The usual case
$\mathcal{S}_k[\widetilde{\rho}_n] = -\ln (4) + f(n)$
is recover as $\gamma\rightarrow 0$.
An analytic expression for $f(n)$ has not yet been found.
However, from a numerical analysis it is obtained that $f(n)$
increases with $n$.
Specifically, we have that
$f(1) \simeq 3.21204$,
$f(2) \simeq 3.60700$ and
$f(3) \simeq 3.75314$.
It is known that
$\lim_{n\rightarrow \infty}f(n) = \ln (8 \pi) + 2(1-c)$,
where $c$ is the Euler-Mascheroni constant.

Hence, the Shannon entropy sum for a particle with
a PDM results
\begin{eqnarray}
\label{eq:Sx+Sk-pdm}
\mathcal{S}_x [{\rho}_n] + \mathcal{S}_k [\widetilde{\rho}_n] =
f(n)-1 + \ln( \sqrt{1 - \gamma^2 a^2}).
\end{eqnarray}
For the usual case the sum
$(\mathcal{S}_x [{\rho}_n] +
\mathcal{S}_k [\widetilde{\rho}_n])|_{\gamma=0}
= f(n) -1$
satisfies the entropic uncertainty relationship
(\ref{eq:entropic-uncertainty-x-k})
once $f(n) -1$ varies from $2.21204$ to $3.06974$.
Thus, for a particle with a PDM within the range $|\gamma a| < 1$
we have that the sum (\ref{eq:Sx+Sk-pdm}) obeys
also the entropic uncertainty relation
(\ref{eq:entropic-uncertainty-x-k}).

From the expressions of
$\mathcal{S}_x[{\rho}_n]$
and $\mathcal{S}_k[\widetilde{\rho}_n]$,
we obtain the following position and wave-vector Shannon length
\numparts
\begin{eqnarray}
\label{eq:L_S-x-space}
L_{\rm S}[{\rho}_n]
&=& \exp (S[{\rho}_n]) = a\, {\exp}(
\mathcal{S}_x[{\rho}_n])
 =  \frac{2L_\gamma}{\rme}\sqrt{1-\gamma^2 a^2},
\\
\label{eq:L_S-k-space}
L_{{\rm S}}[\widetilde{\rho}_n]
&=& \exp ({S}[\widetilde{\rho}_n])
 = \frac{1}{a}\, {\exp}(
\mathcal{S}_k[\widetilde{\rho}_n])
 =  \frac{{\exp}({f(n)})}{2L_\gamma}.
\end{eqnarray}
\endnumparts

Recalling the standard Fisher information in the position space
$\{|\hat{x} \rangle \}$
\begin{equation}
\label{eq:standard-FI-position-space}
F[{\rho}_n] = \int_{-\infty}^{+\infty} \frac{1}{\rho_n (x)}
\left[ \frac{\rmd {\rho}_{n}(x)}{\rmd x} \right]^2 \rmd x
= 4 \int_{-\infty}^{+\infty}
\left[ \frac{\rmd {\psi}_{n}(x)}{\rmd x} \right]^2 \rmd x
= \frac{4\langle \hat{p}^2 \rangle}{\hbar^2},
\end{equation}
and using the moment (\ref{eq:expected-value-p^2}), we have
\begin{equation}
\label{eq:F_x}
F [{\rho}_{n}] =
\frac{4 k_{\gamma,n}^2 \gamma a}{(1-\gamma^2 a^2)^2
\,{\rm atanh}(\gamma a)}
\left[ 1 +
\frac{{\rm atanh}^2(\gamma a)}{4\,{\rm atanh}^2(\gamma a)
+(n{\pi})^2}\right].
\end{equation}
The Fisher information for $k$-space is obtained in a similar way.
Writting the wave function (\ref{eq:psi_k-space}) in the form
$\widetilde{\psi}_n (k) = \zeta_n (k) e^{-\rmi \alpha_n (k)}$,
with $\zeta_n (k) = \sqrt{\widetilde{\rho}_n(k)}$ and
$\alpha_n(k)= \frac{1}{2}
[k\gamma^{-1} \ln (1-\gamma^2 a^2) + \pi (n+1)],$
one obtains
\begin{equation}
F[\widetilde{\rho}_n]
	 = \int_{-\infty}^{+\infty} \frac{1}{\widetilde{\rho}_n (k)}
	    \left[ \frac{\rmd \widetilde{\rho}_{n}(k)}{\rmd k} \right]^2 {\rmd}k
     = 4\int_{-\infty}^{+\infty}
		\left[ \frac{\rmd {\zeta}_{n} (k)}{\rmd k} \right]^2 \rmd k
	 = 4\langle \hat{\eta}^2 \rangle,
\end{equation}
The probability distribution in the deformed space is
$\varrho_n(\eta) = |\phi_n(\eta)|^2 =
A_\gamma^2 \sin^2 (k_{\gamma, n} \eta)$
for
$\gamma^{-1}\ln (1-\gamma a) < \eta < \gamma^{-1}\ln (1+\gamma a)$
and $\varrho_n(\eta)  = 0$ otherwise.
Therefore, it is straightforward verify that
\begin{equation}
\label{eq:F_k}
F[\widetilde{\rho}_n] = 4L_\gamma^2 \left[\frac{1}{3}
- \frac{\ln (1 + \gamma a)}{\gamma L_\gamma}
+ \frac{\ln^2 (1 + \gamma a)}{\gamma^2 L_\gamma^2}
- \frac{1}{2 (n{\pi})^2} \right].
\end{equation}
For $\gamma = 0$, equations (\ref{eq:F_x}) and (\ref{eq:F_k})
recover respectively the usual cases
$F[{\rho}_n] = \frac{\pi^2 n^2}{a^2}$ and
$F[\widetilde{\rho}_n] = \frac{4a^2}{3}
\left( 1- \frac{6}{\pi^2 n^2} \right)$,
according to \cite{Lopez-Rosa-2011}. Also, we have
\numparts
\begin{equation}
\label{eq:L_F-x-space}
L_{{\rm F}}[{\rho_n}] = \frac{a}{n \pi}(1-\gamma^2 a^2)
\left(\frac{L_\gamma}{2a} \right)^{\frac{3}{2}}
\left[ 1 +
\frac{{\rm atanh}^2(\gamma a)}{4\,{\rm atanh}^2(\gamma a)
+(n{\pi})^2}\right]^{-\frac{1}{2}}
\end{equation}
and
\begin{equation}
\label{eq:L_F-k-space}
L_{{\rm F}}[{\widetilde{\rho}_n}] =
\frac{1}{2L_\gamma} \left[\frac{1}{3}
- \frac{\ln (1 + \gamma a)}{\gamma L_\gamma}
+ \frac{\ln^2 (1 + \gamma a)}{\gamma^2 L_\gamma^2}
- \frac{1}{2 (n{\pi})^2} \right]^{-\frac{1}{2}}.
\end{equation}
\endnumparts
In agreed with (\ref{eq:disequilibrium}),
the disequilibrium of the particle
with PDM for spaces $x$ and $k$ are
\begin{eqnarray}
\label{eq:D_x}
D[{\rho}_{n}] &=& \frac{3}{4a}
\frac{\gamma^2 a^2}{(1-\gamma^2 a^2)\,{\rm atanh}^2(\gamma a)}
\nonumber \\
&& \times  \frac{4 \pi^4 n^4}{[{\rm atanh}^2(\gamma a)
           + \pi^2 n^2][{\rm atanh}^2(\gamma a) + 4\pi^2 n^2]}
\end{eqnarray}
and
\begin{equation}
\label{eq:D_k}
D[\widetilde{\rho}_{n}]=
\frac{L_\gamma}{6\pi}\left( 1+\frac{15}{2{\pi}^2 n^2}\right).
\end{equation}

\begin{figure}[!htb]
\centering
\begin{minipage}[h]{0.40\linewidth}
\includegraphics[width=\linewidth]{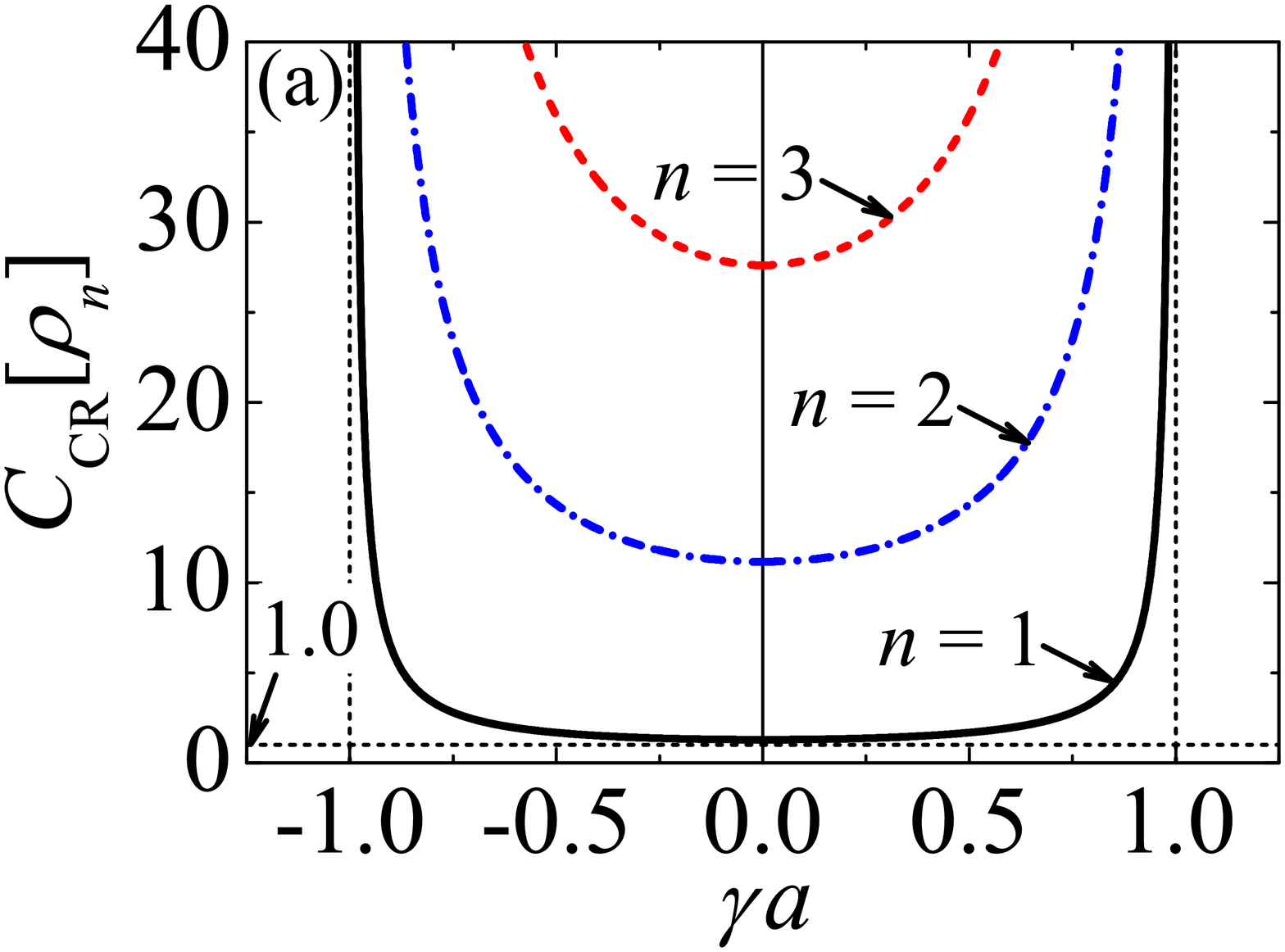}
\end{minipage}
\begin{minipage}[h]{0.40\linewidth}
\includegraphics[width=\linewidth]{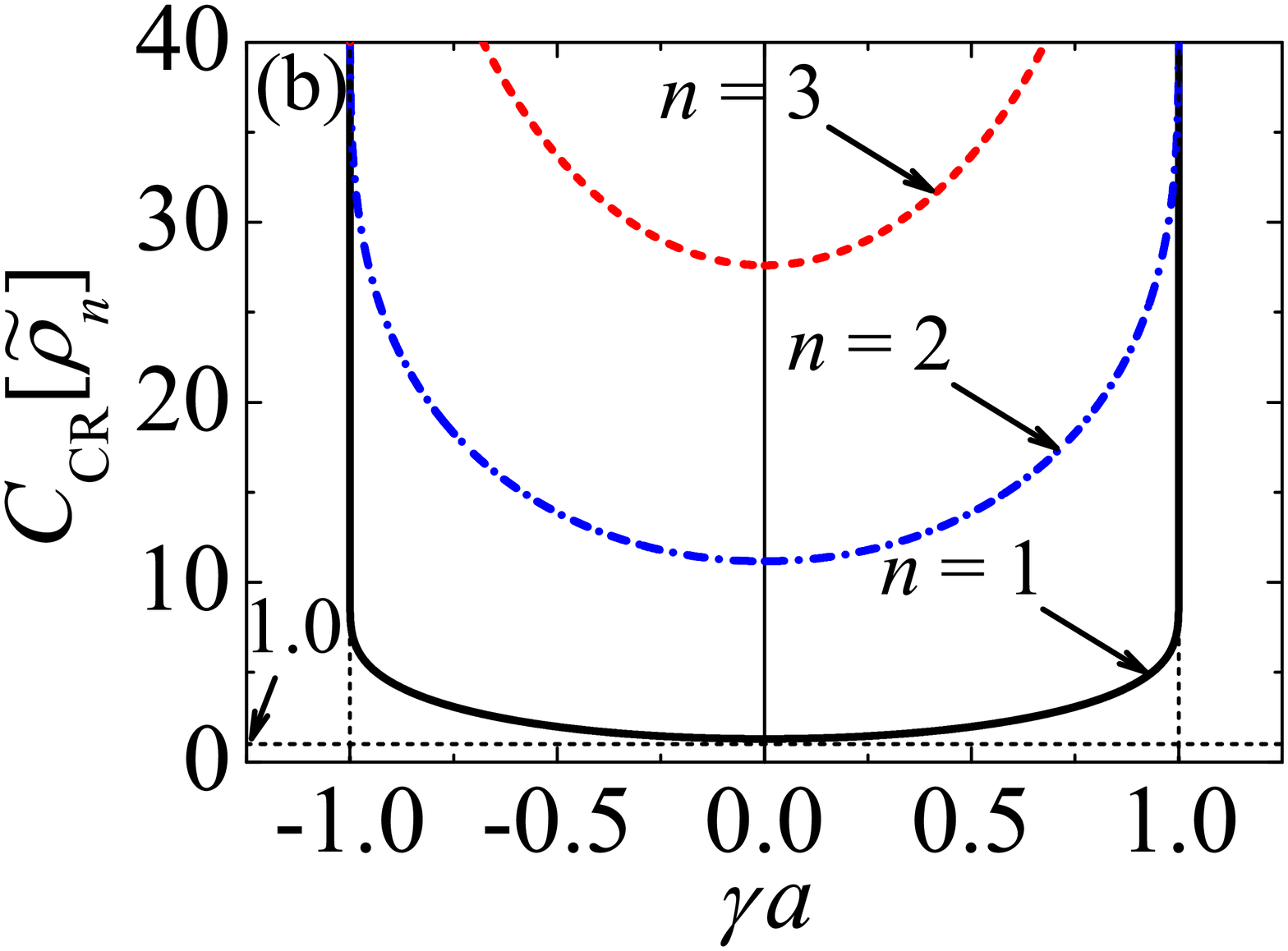}
\end{minipage}\\
\begin{minipage}[h]{0.40\linewidth}
\includegraphics[width=\linewidth]{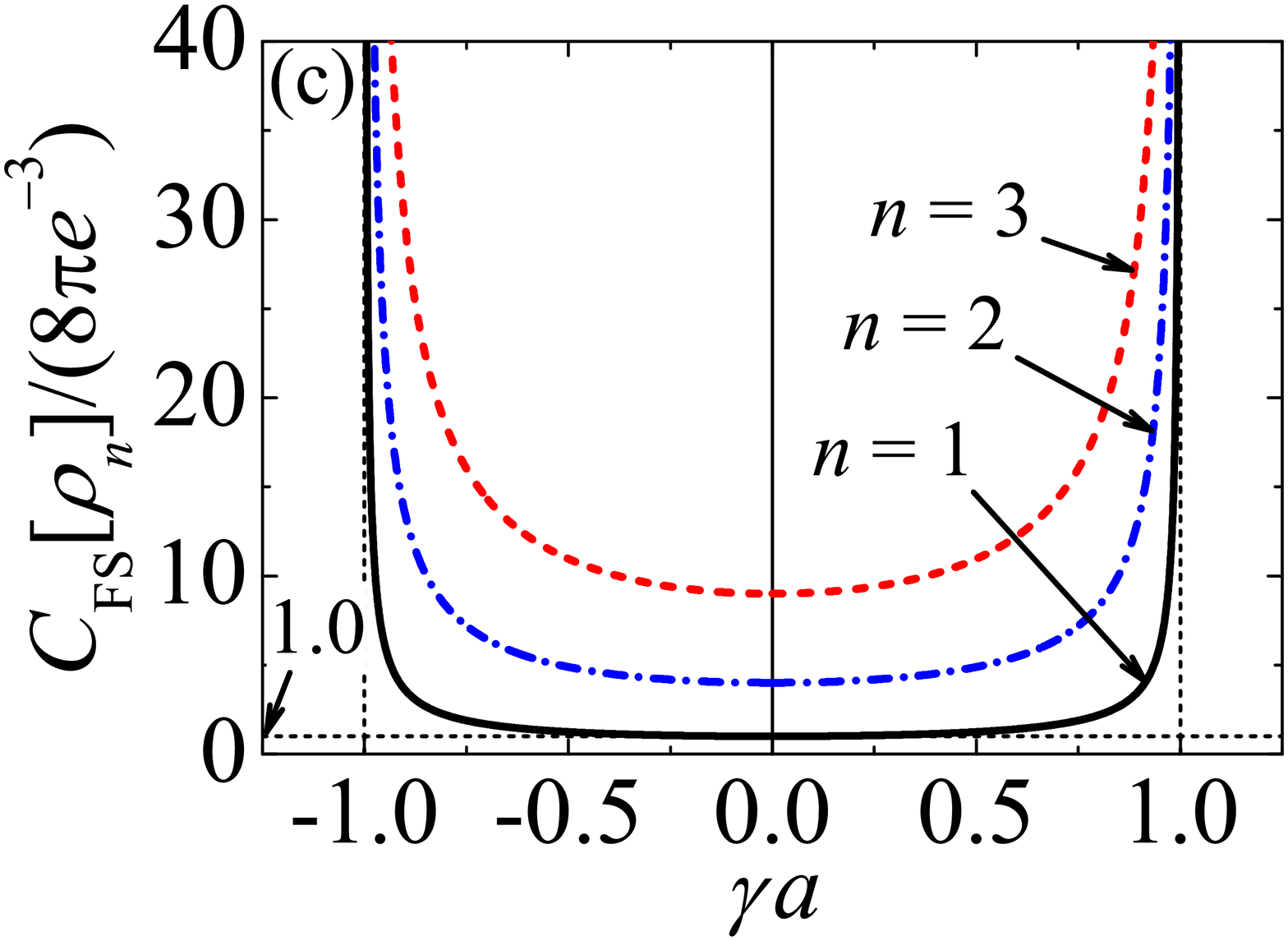}
\end{minipage}
\begin{minipage}[h]{0.40\linewidth}
\includegraphics[width=\linewidth]{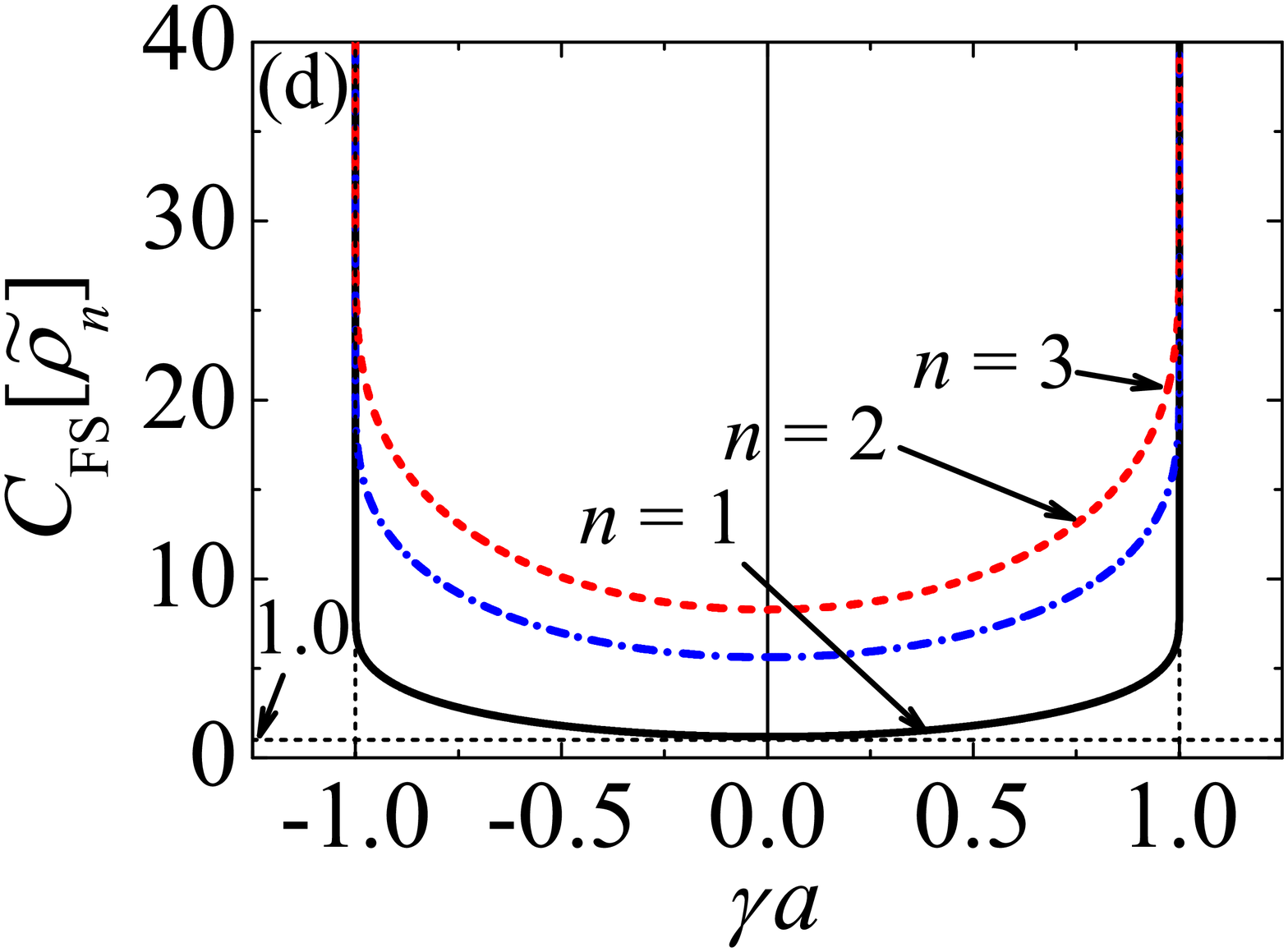}
\end{minipage}\\
\begin{minipage}[h]{0.40\linewidth}
\includegraphics[width=\linewidth]{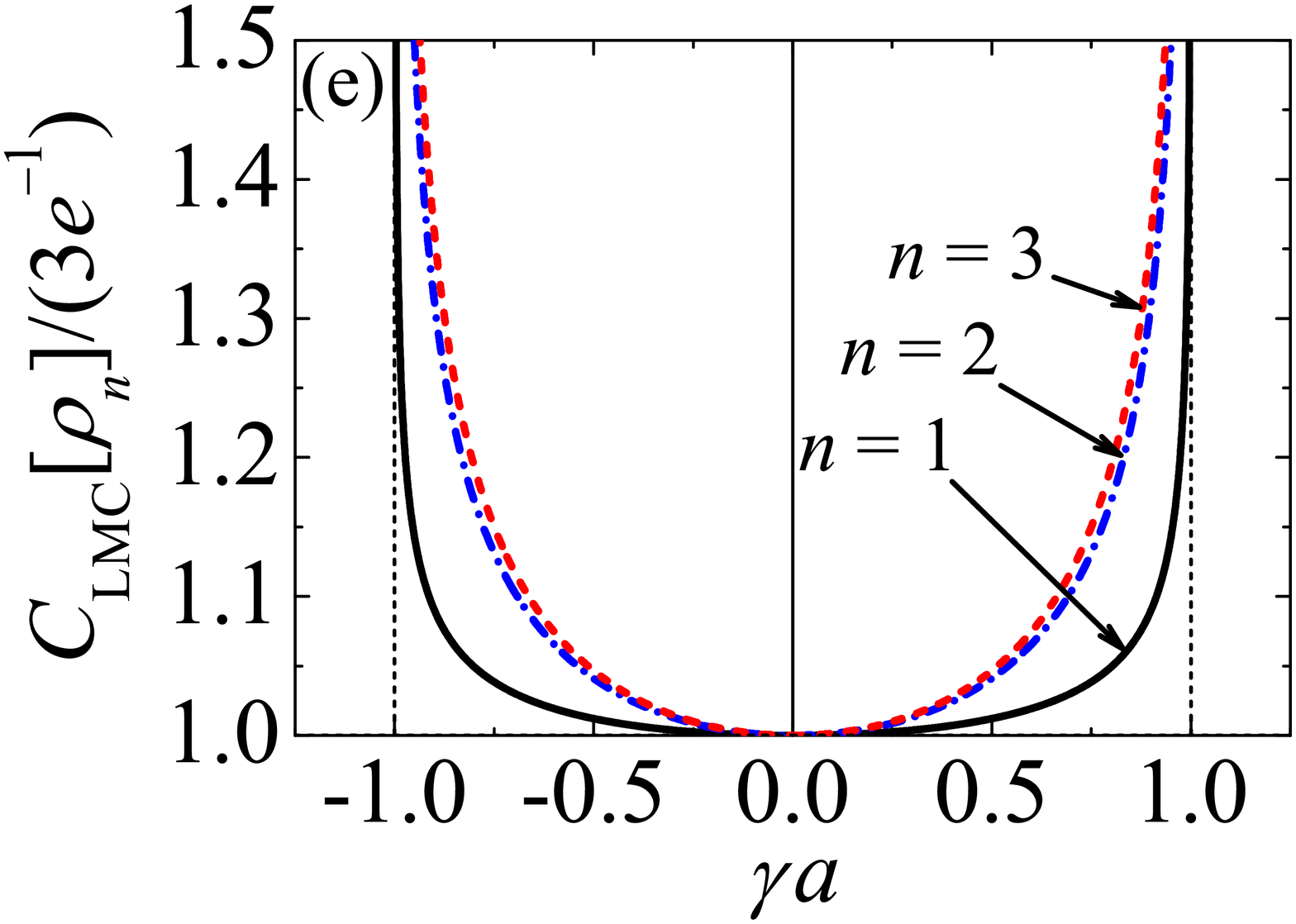}
\end{minipage}
\begin{minipage}[h]{0.40\linewidth}
\includegraphics[width=\linewidth]{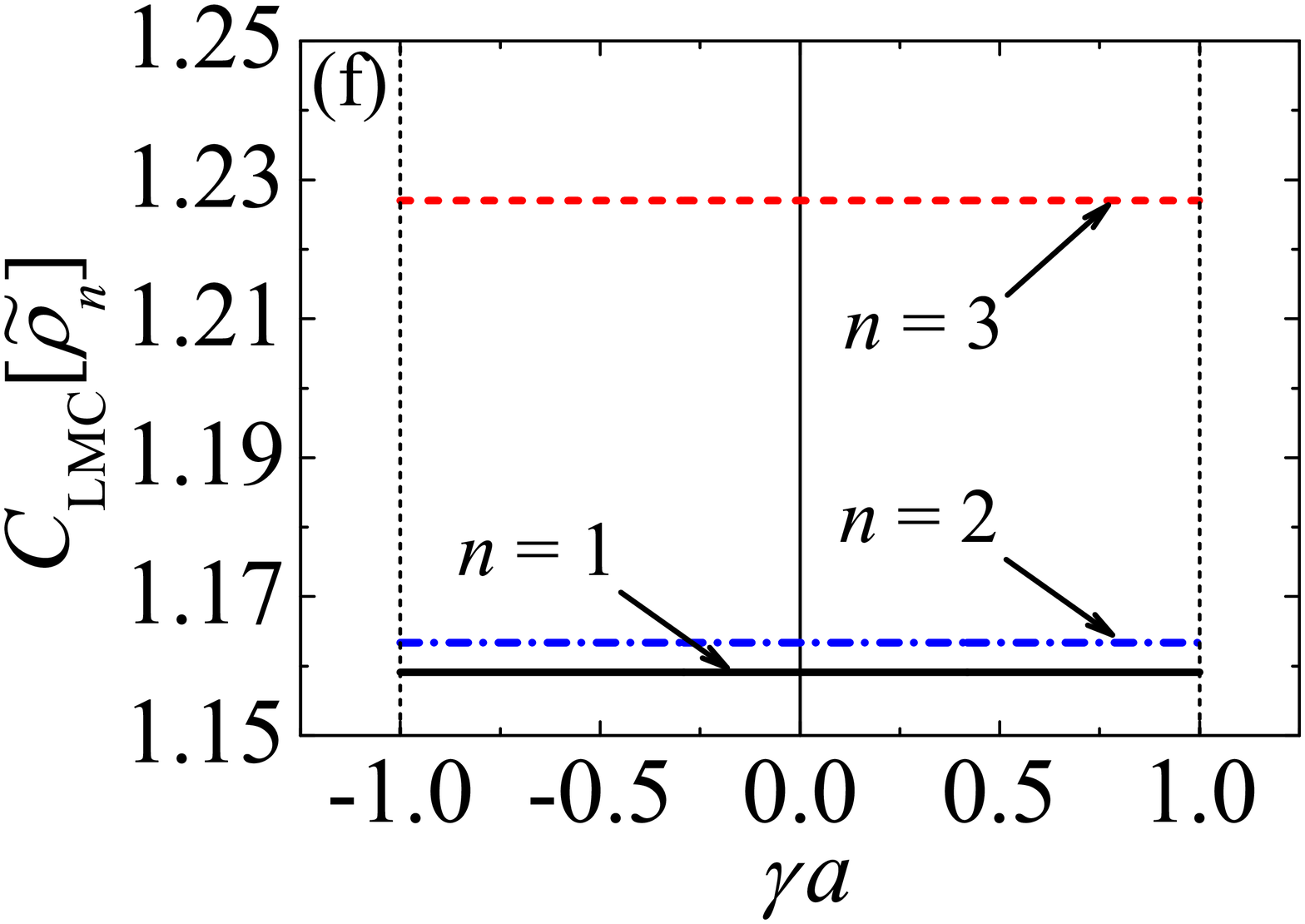}
\end{minipage}\\
\begin{minipage}[h]{0.40\linewidth}
\includegraphics[width=\linewidth]{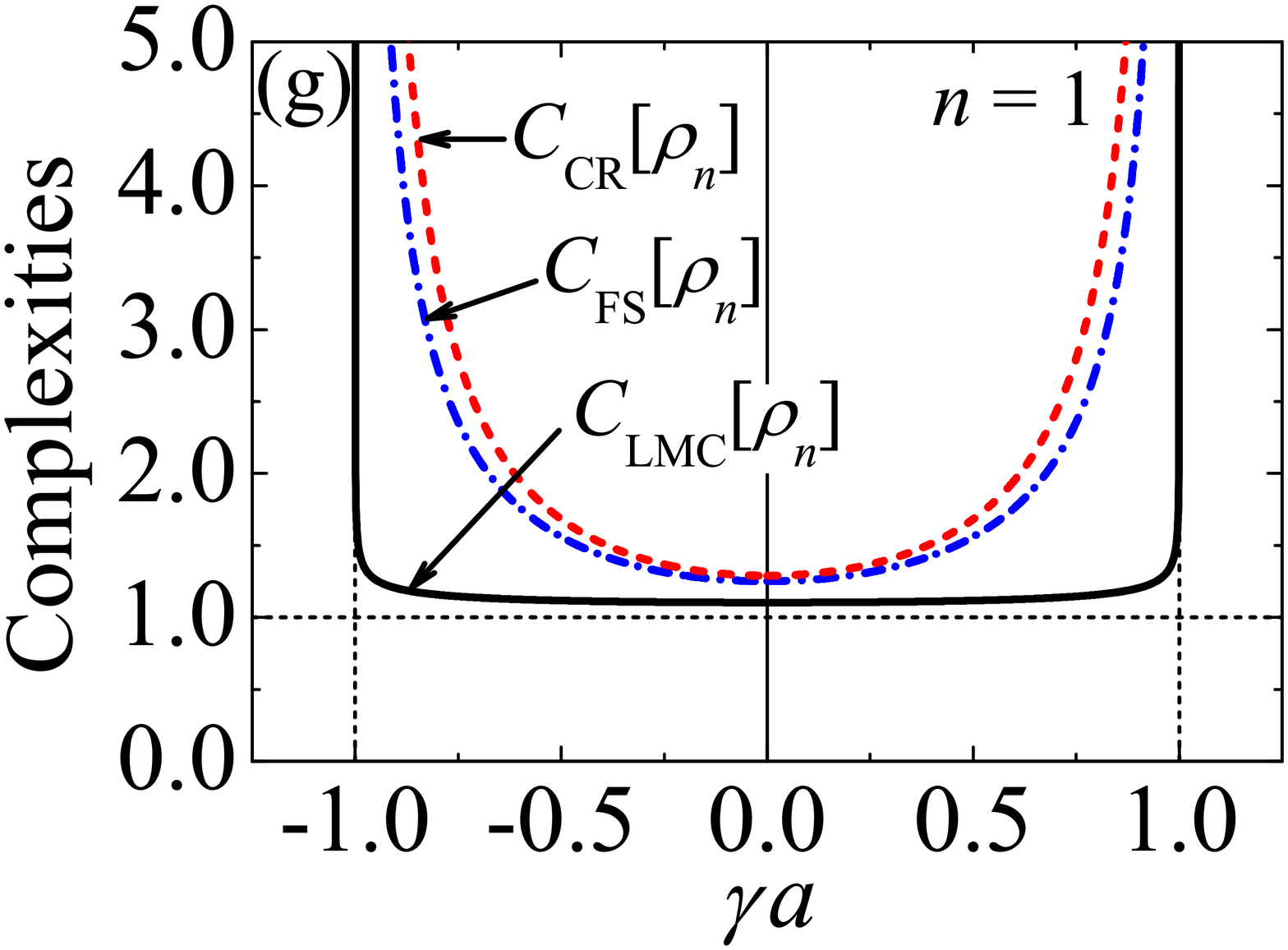}
\end{minipage}
\begin{minipage}[h]{0.40\linewidth}
\includegraphics[width=\linewidth]{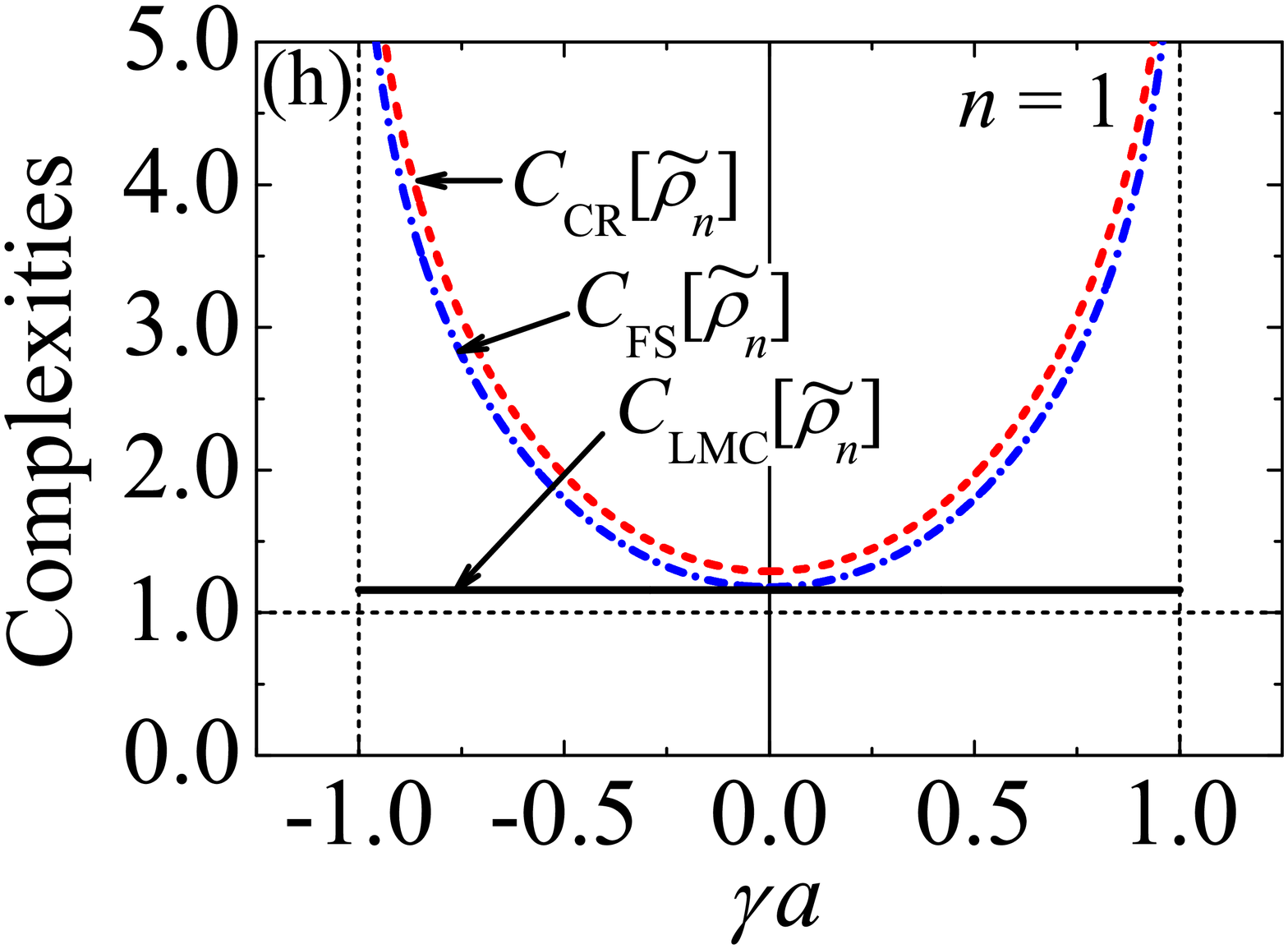}
\end{minipage}\\
\caption{\label{fig:4}
Cram\'{e}r-Rao
[(a) and (b)],
Fisher-Shannon
[(c) and (d)]
and LMC complexities
[(e) and (f)]
for position and wave vector spaces densities
probability
($\rho_n(x)$ and $\widetilde{\rho}_n(k)$)
of the first three eigenstates of a particle with a PDM in
a symmetric infinite square potential well of width
$2a$, in function of the dimensionless deformation parameter $\gamma a$.
The three complexities for the ground state are illustrated for the
$x$ and $k$ spaces in (g) and (h), respectively.
The ordering $C_{{\rm LMC}} < C_{{\rm FS}} < C_{{\rm CR}} $ is
satisfied for the probability distributions in both spaces.
}
\end{figure}

By using (\ref{eq:x-med-quantum}), (\ref{eq:x^2-med-quantum})
and (\ref{eq:F_x}), we get Cram\'er-Rao complexity,
(\ref{eq:CR-complexity}), in position space is
\begin{eqnarray}
\label{eq:CR-PDM-position}
C_{{\rm CR}}[\rho_n] &=&
\left( \frac{ {n\pi \gamma a} }{1-\gamma^2 a^2}\right)^{\! 2}
\frac{1}{{\rm atanh}^5(\gamma a)}
\left[ 1
+ \frac{{\rm atanh}^2(\gamma a)}{4\,{\rm atanh}^2(\gamma a)
+ (n{\pi})^2}
\right]
\nonumber \\
&& \times
\left\{
\frac{{\rm atanh}(\gamma a)(n{\pi})^2}{
4\,{\rm atanh}^2(\gamma a) + (n{\pi})^2}
-\frac{(\gamma a)(n{\pi})^4}{\left[{\rm atanh}^2(\gamma a)
+ (n{\pi})^2\right]^2}
\right\},
\end{eqnarray}
and from (\ref{eq:expected-value-k}),
(\ref{eq:expected-value-k^2}) and (\ref{eq:F_k})
for wave vector space
\begin{eqnarray}
\label{eq:CR-PDM-wavevecktor}
C_{{\rm CR}}[\widetilde{\rho}_n] &=& (2\pi n)^2 \left[\frac{1}{3}
- \frac{\ln (1 + \gamma a)}{2\,{\rm atanh}(\gamma a)}
+ \frac{\ln^2 (1 + \gamma a)}{4\,{\rm atanh}^2(\gamma a)}
- \frac{1}{2 (n{\pi})^2} \right].
\end{eqnarray}
Both complexities (\ref{eq:CR-PDM-position})
and (\ref{eq:CR-PDM-wavevecktor}) recover the usual cases
$C_{{\rm CR}}[{\rho}_n] = C_{{\rm CR}}[\widetilde{\rho}_n]
= 4\left(\frac{\pi^2 n^2}{3} -2\right) > 1$ for $\gamma a = 0$.
On the other hand, the Cram\'er-Rao complexities have different
expressions
in $x$ and $k$ spaces for $\gamma a \neq 0$.
For Rydberg states (i.e., $n\gg 1$) the asymptotic behavior of
(\ref{eq:CR-PDM-position}) and (\ref{eq:CR-PDM-wavevecktor})
are respectively given by
\begin{equation}
C_{{\rm CR}}[\rho_n] \simeq
\left( \frac{ {n\pi \gamma a} }{1-\gamma^2 a^2}\right)^{\! 2}
\frac{1}{{\rm atanh}^4(\gamma a)}
\left[1 - \frac{\gamma a}{{\rm atanh}(\gamma a)} \right],
\end{equation}
\begin{equation}
C_{{\rm CR}}[\widetilde{\rho}_n] \simeq (2\pi n)^2 \left[\frac{1}{3}
- \frac{\ln (1 + \gamma a)}{2\,{\rm atanh}(\gamma a)}
+ \frac{\ln^2 (1 + \gamma a)}{4\,{\rm atanh}^2(\gamma a)}
\right].
\end{equation}
\noindent From (\ref{eq:L_S-x-space}), (\ref{eq:L_S-k-space}),
(\ref{eq:L_F-x-space}) and (\ref{eq:L_F-k-space})
the Fisher-Shannon complexities, (\ref{eq:FS-complexity}),
for $x$ and $k$ spaces are respectively expressed by
\begin{equation}
\label{eq:FS-PDM-position}
C_{{\rm FS}}[\rho_n] = \frac{8\pi n^2}{\rme^3}
\frac{\gamma a}{(1-\gamma^2 a^2) {\rm atanh}(\gamma a)}
\left[ 1 + \frac{{\rm atanh}^2(\gamma a)}{4\,{\rm atanh}^2(\gamma a)
       + (n{\pi})^2}\right],
\end{equation}
and
\begin{equation}
\label{eq:FS-PDM-wavevector}
C_{{\rm FS}}[\widetilde{\rho}_n] =
\frac{\rme^{2{f(n)}-1}}{2\pi}
\left[\frac{1}{3} -
\frac{\ln (1 + \gamma a)}{2\,{\rm atanh}(\gamma a)}
+ \frac{\ln^2 (1 + \gamma a)}{4\,{\rm atanh}^2(\gamma a)}
- \frac{1}{2 (n{\pi})^2} \right].
\end{equation}
For $\gamma a = 0$, one recovers
$C_{{\rm FS}}[\rho_n] = \frac{8\pi n^2}{\rme^3}$ and
$C_{{\rm FS}}[\widetilde{\rho}_n]=
\frac{\rme^{2f(n)-1}}{24\pi} \left( 1 -\frac{6}{\pi^2 n^2} \right).$
The Rydberg states for (\ref{eq:FS-PDM-position})
and (\ref{eq:FS-PDM-wavevector}) become
\numparts
\begin{equation}
\label{eq:FS-PDM-position-n-large}
C_{{\rm FS}}[\rho_n] \simeq \frac{8\pi n^2}{\rme^3}
\frac{\gamma a}{(1-\gamma^2 a^2) {\rm atanh}(\gamma a)},
\end{equation}
\begin{equation}
C_{{\rm FS}}[\widetilde{\rho}_n] \simeq 32\pi \rme^{3-4c}
\left[\frac{1}{3}
- \frac{\ln (1 + \gamma a)}{2\,{\rm atanh}(\gamma a)}
+ \frac{\ln^2 (1 + \gamma a)}{4\,{\rm atanh}^2(\gamma a)} \right].
\end{equation}
\endnumparts
\noindent
The LMC complexities (\ref{eq:LMC-complexity}) are
\begin{equation}
\label{eq:LMC-PDM-position}
C_{{\rm LMC}}[\rho_n] = \frac{3}{\rme}
\frac{\gamma a(1-\gamma^2 a^2)^{-\frac{1}{2}}(4 \pi^4 n^4)}{
{\rm atanh}(\gamma a) [{\rm atanh}^2(\gamma a)
           + \pi^2 n^2][{\rm atanh}^2(\gamma a) + 4\pi^2 n^2]},
\end{equation}
and
\begin{equation}
\label{eq:LMC-PDM-wavecector}
C_{{\rm LMC}}[\widetilde{\rho}_n] =
\frac{\rme^{f(n)}}{12\pi}
\left( 1+\frac{15}{2{\pi}^2 n^2}\right),
\end{equation}
with $C_{{\rm LMC}}[{\rho_n}] =3/\rme$ for $\gamma a = 0$.
Interestingly, $C_{{\rm LMC}}[\widetilde{\rho}_n]$
depends only on $n$ as in the standard case.
Again, considering the Rydberg states, we have
\begin{equation}
C_{{\rm LMC}}[{\rho_n}] \simeq \frac{3}{\rme}
\frac{\gamma a}{\sqrt{1-\gamma^2 a^2} {\rm atanh}(\gamma a)},
\end{equation}
\begin{equation}
C_{{\rm LMC}}[\widetilde{\rho}_n] \simeq \frac{2}{3} \rme^{2(1-c)}.
\end{equation}
\noindent Given a fixed value of the deformation $\gamma a$,
from the expressions
(\ref{eq:CR-PDM-position})-(\ref{eq:LMC-PDM-wavecector})
we have that
$C_{{\rm CR}}[\rho_n]$, $C_{{\rm CR}}[\widetilde{\rho}_n]$,
$C_{{\rm FS}}[\rho_n] \propto n^2$,
for $n\gg1$, which expresses the quadratic dependence with $n$ for
these complexities in the classical limit.
By contrast, when $n\gg1$ we have
$C_{{\rm FS}}[\widetilde{\rho}_n]$,
$C_{{\rm LMC}}[\rho_n]$ and
$C_{{\rm LMC}}[\widetilde{\rho}_n]$
are independent of $n$.
The figure \ref{fig:4} shows the complexities
(\ref{eq:CR-complexity}), (\ref{eq:FS-complexity}) and
(\ref{eq:LMC-complexity}) for distributions
(\ref{eq:rho_n(x)}) and (\ref{eq:rho_n(k)})
of the first three excited states.
For the probability density in position space, the Cram\'er-Rao and
Fisher-Shannon complexities present a similar behavior, while the
LMC complexity is very different from those by exhibiting more
sensibility to the deformation.
Although the entropy densities in position are asymmetric with
respect the axis $x=0$
(figure \ref{fig:3}),
all the complexities turn out symmetric around
$\gamma a=0$ (absence of deformation).
We also observe that the curves have a similar shape
between the Cram\'er-Rao and Fisher-Shannon complexities
with respect to the deformation range of values analysed
$-1 < \gamma a < 1$
(plots (a), (b), (c) and (d) of
the figure \ref{fig:4}),
while the LMC complexity is the more affected by the deformation
(plots (e) and (f) of the figure \ref{fig:4}).
The abrupt variation of the Cram\'er-Rao and Fisher-Shannon
complexities observed from a certain value of $\gamma a$ (which
depends on $n$) indicates the sensibility of the Fisher
information.
The oscillations of the entropy densities are reflected in
the curves of $C_{{\rm CR}}$ and $C_{{\rm FS}}$
(in position and wave vector spaces) that are located from bottom
to up as the quantum number increases.
This does not occurs with the LMC complexity since the curves of
$C_{{\rm LMC}}[\rho_n]$
corresponding to $n=2$ and $n=3$ are almost superposed as well as
$C_{{\rm LMC}}[\widetilde{\rho}_n]$ to $n=1$ and $n=2$.
Plots (e) and (f) express the uncertainty principle between
the position and the wave vector by means of their LMC
complexities. We see that while the LMC's in position space
of the second and the third excited states are almost
superposed the corresponding LMC's curves in wave vector space are
straight lines maximally separated.
For comparison, the plots (g) and (h) illustrate the complexities
of the ground state $n = 1$. It can be observed that in both
$x$ and $k$ spaces the order
$C_{{\rm CR}} > C_{{\rm FS}} > C_{{\rm LMC}} > 1$ is satisfied.

\section{\label{sec:conclusion}
		 Conclusions}

We have studied three main complexity measures on continuous
probability density distributions used in the literature,
the Cram\'{e}r-Rao, the Fisher-Shannon and the LMC ones,
in the context of the deformed
Schr\"{o}dinger equation for PDM systems
proposed originally by Costa Filho {\it et al.}
in reference \cite{CostaFilho-Almeida-Farias-AndradeJr-2011}.
To analyse their behaviors we have
considered some eigenstates of a particle with a PDM and
confined in an infinite potential well.

The effect of the PDM is characterised by the arising of an
asymmetry in the entropy density in position space around the value
$x=0$ (that depends on the functional form of the PDM) and by
regions with negative values, that become
more pronounced as the deformation increases for $\gamma a>0$.
In wave vector space the entropy density is symmetric
around $k=0$, but becomes more compressed when
the deformation parameter increases.

The complexities are symmetric with respect to the
deformation parameter in both position and wave vector spaces.
The Cram\'{e}r-Rao
and Fisher-Shannon complexities
allow to distinguish the eigenstates in the
presence of the deformation, in such a way that they are well
separated (see figure \ref{fig:4}).
In addition, the allowed range of the deformation is bounded
$|\gamma a|\leq 1$,
where the critical values $|\gamma a|=1$ correspond to divergences
of the complexities studied.
Regarding the classical limit, the quadratic dependence
with the quantum number is recovered for the Cram\'{e}r-Rao
and the Fisher-Shannon complexities.
From our study we conclude that in the
context of PDM systems the complexities
in position and wave vector space can provide the same
information, as we see for the Cram\'er-Rao (plots (a) and
(b) of figure \ref{fig:4}) and the
Fisher-Shannon ones (plots (c) and (d) of figure \ref{fig:4}), by
separating  states almost in the same manner.
Moreover, if the complexity curves are superposed (i.e.,
they can not distinguish between
different states as we can see from the plot (e) of figure
\ref{fig:4} for the LMC) the corresponding ones in the
wave vector space must be separated due to the uncertainty
principle. Hence, given a certain complexity,
in practice it is necessary to calculate it in the position
and the wave vector representations
in such a way that, if one of them can not distinguish states
the other will do it according to the uncertainty
principle. This is a feature that has not classical
analogue and then, it can not be used within an exclusively
classical context.

Finally, we consider that the present work gives interesting
features for studying the relationship between the PDM
systems and the information theoretical measures:
abrupt variation of the complexity near to the asymptotic
value of the PDM, erasure of the asymmetry of the
complexity and an entropic density with negative values.
We hope this can be explored with more examples in future
researches.

\ack
I S Gomez acknowledges support received from the National
Institute of Science and Technology for Complex Systems (INCT-SC),
and from the
Conselho Nacional de Desenvolvimento Cient\'ifico e Tecnol\'ogico
(CNPq)  (at Universidade Federal da Bahia), Brazil.

\section*{References}



\end{document}